\documentclass[12pt]{article}
\usepackage[dvips]{graphicx}
\usepackage{subfigure}

\usepackage{epsfig,amsmath,amssymb,verbatim,mathrsfs}

\def\beq{\begin{eqnarray}}
\def\eeq{\end{eqnarray}}
\def\bea{\begin{eqnarray}}
\def\eea{\end{eqnarray}}

\newcommand{\ir}{\text{\tiny IR}}
\newcommand{\uv}{\text{\tiny UV}}

\newcommand{\x}{\text{\tiny X}}

\newcommand{\cft}{\text{\tiny CFT}}
\newcommand{\weak}{\text{\tiny WEAK}}
\newcommand{\eu}{\text{\tiny EU}}

\newcommand{\sm}{\text{\tiny SM}}
\newcommand{\pd}{\text{\tiny PD}}
\newcommand{\hid}{\text{\tiny Hid}}

\newcommand{\dm}{\text{\tiny DM}}
\newcommand{\be}{\begin{equation}}
\newcommand{\ee}{\end{equation}}

\newcommand{\rs}{\text{\tiny RS}}

\begin{document}

\setlength{\baselineskip}{0.2in}


\begin{titlepage}
\noindent
\flushright{Mar 2012}
\vspace{0.2cm}

\begin{center}
  \begin{Large}
    \begin{bf}
Secluded Dark Matter Coupled to a Hidden CFT
     \end{bf}
  \end{Large}
\end{center}

\vspace{0.2cm}

\begin{center}

\begin{large}
{ Benedict von~Harling$^{(1)}$ and Kristian~L.~McDonald$^{(2)}$}\\
\end{large}
\vspace{0.5cm}
  \begin{it}
$(1)$ ARC Centre of Excellence for Particle Physics at the Terascale, School of Physics, University of Melbourne, Victoria 3010,
Australia. \\\vspace{0.5cm}
$(2)$ Max-Planck-Institut f\"ur Kernphysik,\\
 Saupfercheckweg 1, 69117 Heidelberg, Germany.\\\vspace{0.5cm}
\vspace{0.3cm}
Email: bvo@unimelb.edu.au~, kristian.mcdonald@mpi-hd.mpg.de
\end{it}
\vspace{0.5cm}

\end{center}

\begin{abstract}

Models of secluded dark matter offer a variant on the standard WIMP picture and can modify our expectations for hidden sector phenomenology and detection. In this work we extend a minimal model of secluded dark matter, comprised of a $U(1)'$-charged dark matter candidate, to include a confining hidden-sector CFT. This provides a technically natural explanation for the hierarchically small mediator-scale, with hidden-sector confinement generating $m_{\gamma'}\ne0$. Furthermore, the thermal history of the universe can differ markedly from the WIMP picture due to ($i$) new annihilation channels, ($ii$) a (potentially) large number of hidden-sector degrees of freedom, and ($iii$) a hidden-sector phase transition at temperatures $T\ll M_{\dm}$ after freeze out. The mediator allows both the dark matter and the Standard Model to communicate with the CFT, thus modifying the low-energy phenomenology and cosmic-ray signals from the secluded sector.

\end{abstract}

\vspace{1cm}

\end{titlepage}

\setcounter{page}{1}


\vfill\eject


\section{Introduction\label{sec:setup}}
The so-called ``WIMP miracle'' suggests that one may account for the dark matter (DM) of the universe by a minimal extension of the Standard Model (SM) --- namely by including a DM candidate whose mass and coupling parameters are typical of those in the electroweak sector. 
However, motivated either by experimental observations or by a desire to explore a wider range of possibilities, a number of authors have considered models with richer dark (or hidden) sectors in recent years. Typically, such models contain additional forces and/or particles, and can have important consequences for the experimental efforts to discover DM. As an example, models of Secluded Dark Matter can include DM candidates that are not directly charged under the SM but instead couple to the visible sector through the exchange of a ``mediator'' ~\cite{Pospelov:2007mp,Finkbeiner:2007kk}. A hidden sector comprised of a DM candidate ($\psi$) that is charged under a weakly-Higgsed $U(1)'$ symmetry provides a simple example. Hidden sector interactions like $\psi\bar{\psi}\leftrightarrow \gamma' h'$ generate a hidden thermal plasma that can be brought into equilibrium with the SM via gauge kinetic mixing~\cite{Holdom:1985ag} between  $U(1)'$ and hypercharge. Given our current ignorance regarding the dark sector of the universe, it seems prudent to keep such possibilities in mind. Indeed, models of this general type, in which the DM is charged under some hidden sector symmetry (or symmetries) and annihilates into hidden-sector fields rather than SM fields, have received much attention in connection with experiments like INTEGRAL~\cite{Knodlseder:2003sv,Weidenspointner:2006nua}, PAMELA~\cite{Adriani:2008zr}, $Fermi$~\cite{Abdo:2009zk,Ackermann:2010ij}, ATIC~\cite{:2008zzr,Panov:2011zw}, and others. 

\begin{figure}[t]
\begin{center}
        \includegraphics[width = 0.8\textwidth]{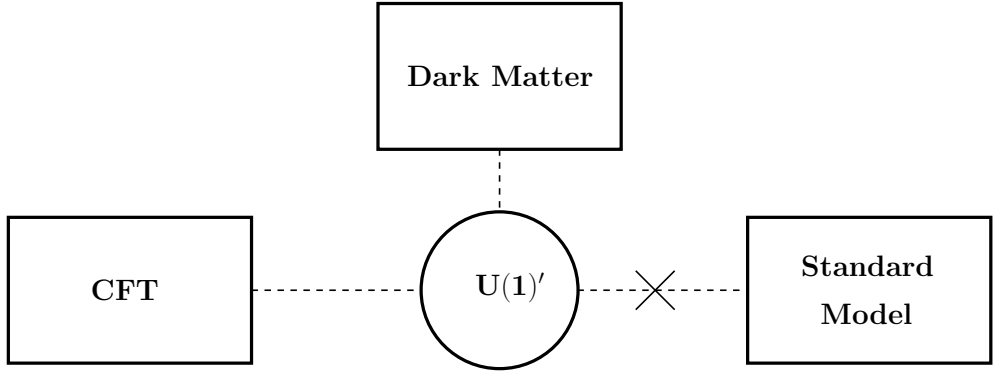}
\end{center}
\caption{Sketch of the Secluded Dark Matter model. The dark matter is
  charged under a gauged $U(1)'$ symmetry of a secluded CFT. Communication between the Standard Model and
  the hidden sector is mediated by $\gamma'$ and proceeds through the kinetic mixing
  with hypercharge (represented by the cross).}
\label{sketch_sdm}
\end{figure}

In this work we consider a variant model of Secluded Dark Matter in which a $U(1)'$-charged 
DM candidate communicates with a strongly-coupled hidden-sector CFT. When the CFT develops a mass gap
in the infrared (IR), this extension allows one to generate a hierarchically small mediator-scale in a technically natural (i.e.~radiatively stable) way. Furthermore, such models contain novel features that can have important consequences for the properties and observability of both DM and the hidden sector more generally. A sketch of the model is given in Figure~\ref{sketch_sdm}. The hidden sector consists of a DM candidate $\psi$ that is charged under a weakly-gauged global $U(1)'$ symmetry of  a hidden CFT. The hidden sector  communicates with the SM via gauge kinetic mixing between $U(1)'$ and hypercharge. Conformal symmetry in the hidden sector  is broken in the IR, with the mass gap set by the confinement scale $\Lambda_{\ir}$. The $U(1)'$ symmetry is also broken when the CFT confines, giving $m_{\gamma'}\sim \Lambda_{\ir}$ and providing a (technically) natural explanation for the hierarchically small value of $m_{\gamma'}$. Thus, we consider a weakly-gauged hidden $U(1)'$ symmetry that is broken by strong dynamics.  Furthermore, we consider the regime $M_{\dm}\gg \Lambda_{\ir}$, so  the hierarchy $m_{\gamma'}\ll M_{\dm}$ is also generated naturally. In the confined phase the CFT contains a tower of spin-one composites $\rho_n$, labeled by the integer $n=1,2,\dots$, with discrete masses $m_{\rho_n}\sim n\times \Lambda_{\ir}$ for $n=\mathcal{O}(1)$, and forming an effective continuum for $n\gg1$.  These composite states mix with $\gamma'$ and thus allow the DM to communicate with the CFT.

As we will show, in order to study this general framework one need not specify the precise details of the hidden CFT. However, one can think of the hidden sector as having the gauge structure $SU(N)'\times U(1)'$, where the $SU(N)'$ sector is conformal for energies $E\gtrsim \Lambda_{\ir}$ and has large $N$. Note the analogy with the low-energy $SU(N_c)\times U(1)$ gauge symmetry of the SM (with $N_c=3$);  one can think of the DM candidate $\psi$ as being analogous to the electron, which is itself charged under a weakly-gauged global symmetry of the strongly interacting QCD sector. Indeed, the mixing between the composites $\rho_n$ and $\gamma'$ mentioned above is analogous to the mixing between the $\rho$ meson and the photon in the SM. Of course, the QCD sector is not conformal above the confinement scale $\Lambda_{QCD}$, but the analogy remains useful.

A detailed discussion of the thermal history of the hidden sector appears in the text. However, for a thermally generated DM abundance the key points are readily summarized. There are two main processes that keep the hidden sector in equilibrium at temperatures $T\sim M_{\dm}$. The $s$-channel process $\psi\bar{\psi}\rightarrow CFT$  proceeds through an off-shell $\gamma'$ and has the cross section\footnote{The number of flavors for the $U(1)'$ sector is $\sim N_f\times N$, where $N_f$ is the number of flavors in the $SU(N)'$ sector. For $N_f\sim \mathcal{O}(1)$ and $N\gg 1$ this gives $\sim N$ flavors for the $U(1)'$ sector so $SU(N)'$ color acts as $U(1)'$ flavor.}
\bea
\sigma_s \ \sim\ N\times \frac{e'^{4}(\sqrt{s})}{M_{\dm}^2},
\eea
where $e'(\sqrt{s})$ is the running $U(1)'$ coupling constant, evaluated at the center-of-mass energy $\sqrt{s}\sim2M_{\dm}$ for $T\sim M_{\dm}$, and we have used the fact that $M_{\dm}\gg m_{\gamma'}$. This process is somewhat analogous to $e^+e^-\rightarrow \rho$ in the SM, which proceeds through a virtual photon. In the SM this process can only proceed via an on-shell $\rho$ meson  for the specific center-of-mass energy $\sqrt{s}\simeq m_\rho$. However, provided $M_{\dm}/ \Lambda_{\ir}\gg 1$, the hidden-sector counterpart $\psi\bar{\psi}\rightarrow CFT$ can always occur because the $SU(N)'$ sector is in the conformal phase and a spin-one state with mass $ m\simeq 2M_{\dm}$ is always present. 

\begin{figure}[t]
\begin{center}
        \includegraphics[width = 0.4\textwidth]{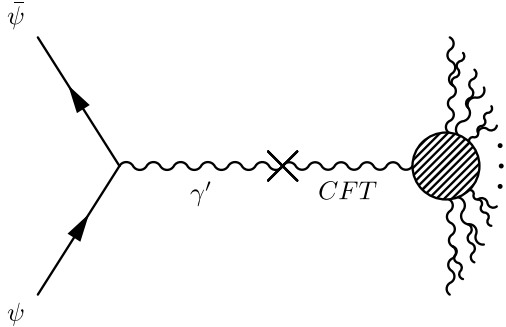}
\end{center}
\caption{$s$-Channel Annihilations: Dark matter
  annihilates into a spin-one state of the CFT, via mixing with the hidden photon, $\gamma'$. Hidden sector showering, represented by the blob, produces a collection of $\sim (M_{\dm}/\Lambda_{\ir})$ light composite states (called $\rho_n$ in the text) and $\gamma'$'s, all with mass $\sim\Lambda_{\ir}$. Though not shown, these light states decay to Standard Model fields through kinetic mixing with hypercharge. }
\label{s_channel_cft}
\end{figure}

The dominant $t$-channel process is $\psi\bar{\psi}\rightarrow 2\gamma'$ and has the standard cross section $\sigma_t \sim  e'^{4}/M_{\dm}^2$, thus giving the simple relation
\bea
\frac{\sigma_s\ }{\sigma_t\ }\ \sim\ N.\label{cross_ratio_4D}
\eea
The annihilation cross section is therefore dominated by the $s$-channel and can be parametrically larger than the standard WIMP value ($\sigma \sim e'^4/M_{\dm}^2$) for large-$N$ theories. This modifies the relationship between the DM mass and coupling required to produce the observed relic abundance.  A full determination of this relationship requires one to account for the presence of the CFT modes in the early universe. The number of degrees of freedom in the thermal plasma is given by $g_*= g_{\sm}+g_{\hid}$, where the hidden sector has $g_{\hid}\sim g_{\cft}\sim \mathcal{O}(N^2)$ degrees of freedom. Clearly, in the large-$N$ limit one can obtain $g_{\cft}/ g_{\sm}\gtrsim\mathcal{O}(1)$ and the effects of the hidden sector can be important. In addition the phase transition from the conformal regime to the confined phase at $T\sim \Lambda_{\ir}$ has further implications for the DM relic abundance.

This scenario clearly differs from the standard WIMP approach and leads to a different picture for the dark sector of the universe. For example, an upper bound on $M_{\dm}$ can be obtained for a standard thermal WIMP (with a given fixed annihilation cross section $\sigma \sim e'^4/M_{\dm}^2$) by the demand that the coupling $e'$ remain within the perturbative regime. In the large-$N$ limit one can (naively) alleviate the upper bound on $M_{\dm}$ by a factor $\sim \sqrt{N}$. In addition, the requisite DM annihilation cross section is modified when the hidden sector contains $g_\hid\gtrsim g_\sm$ degrees of freedom.

The expected signals from DM annihilation and/or scattering can also be modified, with important consequences for present-day experiments. Let us briefly summarize some of the key points. When both the $s$- and $t$-channel annihilations are available in the present-day the $s$-channel is expected to dominate. In this case the annihilation of $\psi\bar{\psi}$ into the CFT is followed by a hidden-sector cascade, ultimately producing a collection of $\sim M_{\dm}/\Lambda_{\ir}$ light states. These consist of the composites\footnote{Technically, the composite spectrum will also contain spin-two modes, so that $\rho_n$ denotes generic composite states. Also, some composites will decay to $\gamma'$ due to the aforementioned kinetic mixing. } $\rho_n$ and the hidden-photon $\gamma'$, all with mass and energy on the order of the confinement scale $\Lambda_{\ir}$. As mentioned already, the modes $\rho_n$ mix with $\gamma'$, which itself mixes with SM hypercharge. In the absence of hidden-sector states with mass $m\ll\Lambda_{\ir}$, the only available decay-channels for the lightest hidden-sector states are to SM fields. Thus, a typical $s$-channel annihilation process can be denoted as
\bea
\psi\ \bar{\psi}\ \longrightarrow\ CFT\ \longrightarrow\  (M_{\dm}/\Lambda_{\ir})\times (\rho_n+\gamma')\ \longrightarrow\ (M_{\dm}/\Lambda_{\ir})\times f\bar{f}\ ,
\eea
where $f$ denotes a light SM field with mass $m_f<\Lambda_{\ir}$. The result is a large number ($\sim M_{\dm}/\Lambda_{\ir}$) of light SM fields with energies of order $\Lambda_{\ir}$.  Such a signal could be quite interesting, though certainly non-standard. The first two steps of this annihilation process are shown in Figure~\ref{s_channel_cft}.

On the other hand, if the DM $\psi$ is split into two Majorana fermions $\psi_{1,2}$ by some small splitting, $\Delta M_{\dm}\ll M_{\dm}$, the present-day abundance is dominated by the lighter state $\psi_1$. The $s$-channel predominantly couples $\psi_{1,2}$ together, $\psi_1\psi_2\rightarrow CFT$, and absent an abundance of $\psi_2$,  is strongly suppressed in the present-day. However, this channel remains present in the early universe at temperatures $T\gtrsim  M_{\dm}$, and therefore affects the thermal history of the DM abundance. In this case present-day annihilations are dominated by the $t$-channel process $\psi_1\psi_1\rightarrow 2\gamma'$, resulting in boosted $\gamma'$ production, with additional sub-dominant production of hidden-sector states, $\psi_1\psi_1\rightarrow \rho_n \rho_{n'}$, due to the $\gamma'-\rho_n$ mixing. The decay of $\gamma'$ to the SM creates boosted SM fields, similar to recent models\footnote{The split-fermion case, with  $M_{\dm}\sim$~TeV and $\Lambda_\ir\sim$~GeV, provides a concrete example of a model with ``new irrelevant annihilation channels''  in accordance with the definitions of Ref.~\cite{Slatyer:2011kg}.} of DM~\cite{ArkaniHamed:2008qn}. The CFT states undergo some showering before producing SM fields. In either case, the precise spectrum of final-state SM fields depends on the ratio $M_{\dm}/\Lambda_\ir$.

In this work we construct a class of models with Secluded Dark Matter coupled to a hidden CFT, and study their viability as theories of DM. The thermal history of the DM can be modified significantly by the new annihilation channels, the potentially large number of hidden-sector degrees of freedom, and the occurrence of a hidden-sector phase transition at late temperatures $T\sim\Lambda_\ir\ll M_{\dm}$. We find that viable models of DM exist and detail the parameter space that is consistent with the demands of thermal DM production and low-energy constraints. Interestingly, we show that the viable parameter space admits kinetic mixing between hypercharge and the light hidden-sector vectors ($\gamma'$ and $\rho_n$) that is large enough to produce signals at ongoing low-energy experiments. Thus, in this framework, a host of interesting low-energy signals are compatible with successful production of the DM abundance. As we will show, this statement holds for a range of hidden-sector confinement scales $\Lambda_\ir$.

As evidenced by the above discussion, key features of our Secluded
Dark Matter framework stem from the postulated strongly-interacting
conformal sector. Performing reliable calculations in such a model is,
in general, not possible due to the breakdown of perturbation
theory. However, via the AdS/CFT correspondence, classes of
strongly-coupled  conformal models in 4D are dual to weakly-coupled 5D
models on a slice of $AdS_5$. Thus we can circumvent the anticipated
problems with strong-coupling by employing the AdS/CFT dictionary to
construct a weakly-coupled (i.e.~calculable) dual model in warped
space. This is the approach that we adopt: We construct an
explicit 5D model whose dual theory contains a hidden CFT with a
weakly-gauged global $U(1)'$ symmetry. If the dual CFT is an $SU(N)'$
theory with large-$N$, as with the best known applications of the
correspondence, the resulting picture is precisely that of a hidden
sector with  $SU(N)'\times U(1)' $ gauge symmetry. The dual framework
allows us to explicitly calculate many important quantities for the
analysis, and provides a convenient tool for the study of a
strongly-interacting hidden sector.

The layout of the paper is as follows. In Section~\ref{sec:first_model} the weakly-coupled dual 5D theory is presented. The coupling between the SM and the hidden sector is considered in Section~\ref{sec:hidden_kk_mode_pheno_one_throat}, and the decay properties and phenomenology of the light hidden-sector states are also detailed. Communication between the DM and the hidden CFT is discussed in 
Section~\ref{sec:DAannihilations} and the dominant DM annihilation channels are analyzed. With these ingredients, we consider the early universe cosmology in Section~\ref{sec:cos} and determine the conditions under which a successful thermal abundance of DM is obtained. We elucidate the expected modifications to the cosmic ray signal in Section~\ref{sec:cosmi_ray}, providing some general discussion and commenting on high-energy cosmic lepton production in particular. A detailed discussion of the model, in terms of the AdS/CFT dictionary, is given in
Section~\ref{ads_cft}. We comment on some modifications and directions for further study in Section~\ref{sec:comm}, and draw conclusions in Section~\ref{sec:conc}. More discussion regarding
the early universe cosmology and the hidden-sector phase transition
appears in a pair of Appendices. 
\section{A Dual Model of Secluded Dark Matter\label{sec:first_model}}
Our goal in the present work is to implement, and determine the
viability of, a model of Secluded Dark Matter in which the mediator scale is generated by a confining hidden-sector CFT. 
In order to study such a model without encountering the computational complexity of a strongly-coupled hidden sector, we invoke the AdS/CFT correspondence and consider a weakly-coupled warped model that is dual to this 4D picture. The basic ingredients in the 4D model are as follows. It should include a strongly-coupled hidden CFT, with a global $U(1)'$ symmetry that is weakly gauged by a ``fundamental'' photon (in the sense that $\gamma'$ is external to the strongly interacting sector). In addition there should be a fundamental DM candidate $\psi$ that is also charged under $U(1)'$. This allows the DM to couple to the CFT via $\gamma'$ exchange. Also, the model should contain the SM, in the form of fundamental fields, with couplings to the CFT mediated via kinetic mixing between hypercharge and $\gamma'$.

A cursory study of the AdS/CFT dictionary reveals that our proposed model of Secluded Dark Matter is dual to a theory constructed on a slice of $AdS_5$ with a bulk $U(1)$ symmetry. The DM candidate and the SM, being fundamental fields, should be localized on the UV brane, while the IR brane should be of order $\Lambda_{\ir}$ to achieve the specified mass gap. The bulk vector should have a Neumann BC in the UV to ensure the spectrum contains a fundamental vector field (the dual of $\gamma'$) and the bulk $U(1)$ symmetry should be broken in the IR to naturally generate the mediator scale of order $\Lambda_{\ir}$. The UV scale is effectively a free parameter, though consistency requires $\Lambda_{\uv}\gtrsim M_{\dm}$. We will focus on the case where $\Lambda_{\uv}\sim M_{Pl}$, which corresponds to a hidden CFT that is cutoff in the UV at the Planck scale. However, the reader should keep in mind that UV conformal symmetry breaking could in principle occur below this scale. We will provide a more detailed discussion of the relationship between the proposed 4D model of Secluded DM with CFT and the warped 5D model in Section~\ref{ads_cft}. For the most part we shall, up to that point, simply present the 5D model and employ it for our analysis.

\begin{figure}[ttt]
\begin{center}
        \includegraphics[width = 0.4\textwidth]{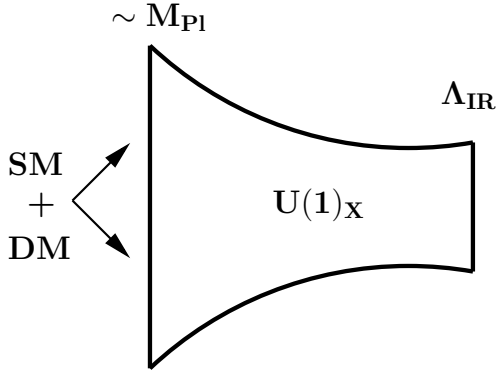}
\end{center}
\caption{A Dual Theory for Secluded Dark Matter: A $U(1)_{\x}$ gauge
  symmetry propagates in a hidden warped-space with IR scale $R^{-1}\sim\Lambda_\ir$, while the Standard Model and dark
  matter are confined to the UV brane. The dark
  matter is charged under $U(1)_{\x}$ and the Standard Model couples to
  the hidden sector via UV-localized gauge kinetic-mixing.}\label{fig:one_throat_setup}
\end{figure}

The basic 5D model is shown in
Figure~\ref{fig:one_throat_setup}. It consists of a
hidden warped-space with IR scale $\Lambda_{\ir}$, UV scale $\sim M_{Pl}$, and a bulk
$U(1)_{\x}$ gauge symmetry.\footnote{We label the 5D $U(1)$ symmetry as $U(1)_{\x}$ and reserve the labels $U(1)'$ and $\gamma'$ for the 4D fields in the dual theory. Similarly we reserve the notation $\rho_n$ for the 4D composite spin-one fields related to the KK vectors $X_n^\mu$ (of the bulk 5D vector $X_M$) by duality.} The
warped space is a slice of $AdS_5$ with metric~\cite{Randall:1999ee}
\beq
ds^2= \frac{1}{(kz)^2}(\eta_{\mu\nu}dx^{\mu}dx^{\nu} -
dz^2)= G_{MN} dx^{M}dx^{N},
\label{bulkmetric}
\eeq
where $z \in [k^{-1},\,R]$ labels the extra dimension, $k$ is the curvature, and $\mu,\nu$
($M,N$) are the 4D (5D) Lorentz indices. The IR brane is located at $z=R\sim \Lambda_{\ir}^{-1}$. The standard calculation of
the effective 4D Planck mass
gives $M_{Pl}^2= (M_*^3/k)\{1-(kR)^{-2}\}$, where $M_*$ is the 5D gravity scale.
The metric is sourced by a bulk cosmological constant and two branes
with non-zero tensions, the latter being located at the ends of the
space. The length of the space is readily stabilized~\cite{Goldberger:1999uk} so no
tuning is required to realize the background (beyond the usual tuning
of the 4D cosmological constant). Ref.~\cite{Agashe:2007zd} argues that higher order terms in the gravitational description of RS models can be safely ignored for $k/M_*<[3\pi^3/(5\sqrt{5})]^{1/2}\sim 3$. We therefore restrict our attention to values of $k/M_*\le 1$ and, except where otherwise stated, employ $k/M_*=\mathcal{O}(0.1)$ for numerical examples. Let us note that previous works have considered a light warped hidden sector~\cite{Gripaios:2006dc}.

As can be seen in Fig.~\ref{fig:one_throat_setup}, the SM is localized on
the UV brane, along with the DM candidate (which we discuss momentarily). This differs
from the usual approach in RS models, where the IR scale is of order
TeV in association with the electroweak scale, and the warping
generates the weak/Planck hierarchy. Here the warping generates the $\Lambda_\ir$/Planck hierarchy but does
not explain the weak/Planck
hierarchy. The mechanism underlying the stability of the weak scale is not of particular relevance for our purposes. As noted in~\cite{McDonald:2010fe}, models such as that of Fig.~\ref{fig:one_throat_setup} can mimic
the low-energy effects of more realistic models in which the weak
scale is also realized via warping. Alternatively, one could readily include SUSY (or some other mechanism) to stabilize the weak/Planck hierarchy. However, for $M_{\dm}\lesssim$~TeV the full details of the theory beyond the TeV scale are not important from the perspective of  the low-energy phenomenology of interest here.\footnote{Introducing bulk SUSY would necessitate SUSY partners for the bulk fields. These would modify the fine details of the hidden-sector phenomenology (e.g.~more decay channels available for hidden-sector Kaluza-Klein modes) but would not disrupt the gross picture espoused in Section~\ref{sec:hidden_kk_mode_pheno_one_throat}.} Thus one can think of the model as
providing an
approximate low energy (sub-$M_{\dm}$) description of a more complete
setup. In cases where the
DM becomes hierarchically large compared to the weak scale,
$M_{\dm}/\langle H\rangle\gg 1$, one should include the dynamics
responsible for stabilizing the weak scale. However, provided the DM
annihilates predominantly into the CFT this will not significantly
affect our analysis.\footnote{The most likely modification is if  the
  stabilizing dynamics introduce a number of new thermal degrees of
  freedom at temperatures $T\lesssim M_{\dm}$; as long as
  $g_{new}\lesssim g_\sm$ (or $g_{new}\lesssim g_\cft$ if $g_\cft\gg g_\sm$) the effects should not be marked.}

In addition to the SM, the particle spectrum
contains towers of Kaluza-Klein (KK) vectors and gravitons, with spacing on the order of $\Lambda_{\ir}$. The spectrum of massive KK gravitons 
(which we denote as $h_a$ with integer $a\ge1$) can be found 
by perturbing around the background metric of Eq.~\eqref{bulkmetric}. 
Their masses are related to the IR scale in the standard way~\cite{Davoudiasl:1999jd}
\bea
m_a \simeq \frac{\pi}{R}(a+1/4),\quad a\ge1\ .
\eea
The massless zero-mode $h_0$ is the usual 4D graviton. Note that
the spectrum necessarily contains KK gravitons. Although our primary
objective is to realize a hidden photon with a stable, order $\Lambda_{\ir}$
symmetry breaking scale, the use of warping
automatically implies the existence of light KK gravitons. For energies $E\ll k$ these modes couple very weakly to the UV-localized SM fields and their direct impact on SM particle phenomenology is negligible. However, they do play a role in the phenomenology of the hidden sector, as we discuss in
Section~\ref{sec:hidden_kk_mode_pheno_one_throat}.

In addition to the KK gravitons, the low-energy spectrum contains a tower of KK vectors. Writing the kinetic action for the bulk vector $X_M$ as
\bea
S_{\x} &=& 
- \frac{1}{4}\int d^4x\,dz\;\sqrt{G}\,G^{MA}G^{NB}X_{MN}X_{AB},
\eea
where  $X_{MN}$ is the $U(1)_{\x}$ field strength, we can perform the usual KK expansion.\footnote{We work in unitary gauge with $X_5=0$.} Symmetry breaking in the hidden sector should occur on (or near) the IR brane to generate an $\mathcal{O}(R^{-1})$ mass for the lightest vector. This can be achieved with either an explicit IR-localized Higgs or by imposing a Dirichlet boundary 
condition in the IR, as in the ``Higgsless'' models~\cite{Csaki:2003dt}.  
We focus primarily on the Higgsless case, 
but will discuss some of the differences with an explicit hidden-Higgs in later
sections.  

We expand the 5D gauge field 
into KK modes as
\beq
X_{\mu}(x,z) = \sum_nf_n(z)X^n_{\mu}(x),
\eeq
where the bulk wave functions satisfy~\cite{Davoudiasl:1999tf}
\beq
\left[z^2\partial_z^2 -z\partial_z + z^2m^2_n\right]f_n(z) = 0,\label{vector_EOM}
\eeq
and the following orthogonality condition:
\begin{eqnarray}
\int  \frac{dz}{(kz)}f_n(z)f_m(z)&=&\delta_{nm}.
\end{eqnarray}
The solutions to \eqref{vector_EOM} are given by
\beq
f_n(z) = \frac{(kz)}{N_n}\left\{J_1(m_nz) +\beta_nY_1(m_nz)\right\}, 
\label{wavefunc-gauge}
\eeq
where $J_1/Y_1$ are Bessel functions.
The Neumann boundary condition at $z=k^{-1}$ gives
\beq
\beta_n = -\frac{J_0(m_n/k)}{Y_0(m_n/k)} \simeq 
\frac{\pi}{2}\;\frac{1}{[\log(2k/m_n)-\gamma]},\label{beta_uv_vector}
\eeq
where the last expression holds for $m_n/k\ll 1$ and $\gamma \simeq 0.5778$ 
is the Euler-Mascheroni constant.

The KK masses $m_n$ are fixed by applying the IR-brane
boundary condition, which, in the Higgsless case, is a Dirichlet BC. This gives 
\beq
\beta_n=-\frac{J_1(m_n R)}{Y_1(m_n R)}\ ,
\label{dirichletfreq}
\eeq
and the KK masses follow from equating this expression with Eq.~\eqref{beta_uv_vector}. For $n$ greater than a few the masses are well approximated by 
\beq
m_n \simeq \frac{\pi}{R}(n+1/4),
\eeq
and are of order $n\times\Lambda_{\ir}$ for $R^{-1}\sim\Lambda_{\ir}$. The normalization factors 
$N_n$ are
\beq
N_n^{-1} \simeq \frac{1}{(n+ 1/4)^{1/2}}\frac{m_n}{\sqrt{k}}.
\eeq
The mass of the lightest mode,
which we label as ``0'', is mildly suppressed relative to the IR scale,
\beq
m_0 \simeq \frac{1}{R}\sqrt{\frac{2}{\log(2kR)-\gamma}},
\eeq
and the wavefunction for this mode is given by Eq.~\eqref{wavefunc-gauge} with
$N_0^{-1} \sim \sqrt{2/kR^2}$.

We consider DM in the form of a UV-localized SM-singlet Dirac fermion
$\psi$, with non-zero $U(1)_{\x}$ charge $Q_x=+1$. The UV-localized DM will couple to the tower of KK vectors and thus the interactions and properties of the KK vectors play an important role in determining the properties of the DM.  It is, therefore, important to understand the decay properties and
phenomenology of the KK vectors. We now turn our attention to this matter and will subsequently consider the DM annihilation channels in detail.
\section{Hidden KK Phenomenology\label{sec:hidden_kk_mode_pheno_one_throat}}
The DM candidate $\psi$ is secluded from the SM. However, the most general action consistent with the symmetries of the theory contains a UV-localized gauge kinetic-mixing term between hypercharge and $U(1)_{\x}$. This mixing plays an important role in the model and provides the dominant coupling between the SM and the hidden sector. We write the UV-localized kinetic-mixing action as
\bea
S &\supset& 
- \frac{\epsilon_*}{2\sqrt{M_*}}
\int_{UV}\!\!d^4x\;\sqrt{-g}\,g^{\mu\nu}g^{\alpha\beta}
B_{\mu\alpha}X_{\nu\beta},
\eea
where $g_{\mu\nu}$ is the induced metric and  $B_{\mu\nu}$ is the SM hypercharge field strength. We assume perturbative values of the
dimensionless parameter $\epsilon_*$, which controls the strength of mixing
between the two sectors, and can be generated by, e.g., integrating out a UV-localized Dirac 
fermion charged under both Abelian factors. The origin of this term is not important as far as the low-energy phenomenology is concerned; the term is consistent with the symmetries of the model and may be included.
 
The kinetic-mixing operator
induces mixing between SM hypercharge and the full tower of KK vectors~\cite{McDonald:2010iq,McDonald:2010fe}. Using the KK expansion for $X^\mu$, we can write the mixing Lagrangian as 
\beq
\mathscr{L}_{eff} \supset -\frac{1}{2}\sum_{n}\epsilon_nX^{\mu\nu}_n
(c_WF_{\mu\nu}-s_WZ_{\mu\nu}),
\eeq
where $s_W$ and $c_W$ refer to the weak mixing angle. The effective
mixing parameter for the $n$-th mode, $\epsilon_n$, is
\beq
\epsilon_0 &=& \frac{\epsilon_*\,f_0(k^{-1})}{M_*^{1/2}} \simeq 
-\epsilon_*\left(\frac{k}{M_*}\right)^{1/2}\frac{1}{\sqrt{\log(2k/m_0)-\gamma}}\ ,
\nonumber\\
\epsilon_n &=& \frac{\epsilon_*\,f_n(k^{-1})}{M_*^{1/2}}
\simeq
\,
-\epsilon_*\left(\frac{k}{M_*}\right)^{1/2}\frac{1}{[\log(2k/m_n)-\gamma]}\,(n+1/4)^{-1/2}\ ,
~~~{n \geq 1}.
\label{epsilonn}
\eeq
For modes with $n>0$ and $m_n\ll k$, the mixing can be approximately related to that of the zero mode:
\bea
\epsilon_n\simeq \frac{\epsilon_0}{\sqrt{n+1/4}}\times \frac{1}{\sqrt{\log(2k/m_n-\gamma)}} \quad \quad\mathrm{for}\  n>0 \, .\label{kk_mixing_relation}
\eea 
Note that the mixing parameter for the higher KK modes is mildly suppressed relative to that of the zero mode, thus suppressing their coupling to the SM. As an example, for $R^{-1}\sim$~GeV one has $\epsilon_n \simeq \epsilon_0/(6\sqrt{n})$ for $k\sim M_{Pl}$.

The DM can annihilate into kinematically available KK vectors (see Section~\ref{sec:DAannihilations}) and a complete analysis of the resulting signals requires one to understand the decay properties of the hidden KK vectors (and gravitons). A detailed analysis of these decays was undertaken
in Ref.~\cite{McDonald:2010fe}. For completeness, we summarize the relevant details of that
work in what follows, and add some additional results of relevance
for the present analysis. Readers that are primarily interested in the gross details of
our DM model can proceed to the next sections.

Due to the kinetic mixing with SM
hypercharge, the KK vectors can decay to SM fields. 
These decays require a kinetic mixing insertion and
the corresponding widths go like
$\epsilon_n^2$. Given the bound of $\epsilon_0\lesssim 10^{-3}$ obtained in Ref.~\cite{McDonald:2010fe} (which we discuss below), and the relationship \eqref{kk_mixing_relation} between $\epsilon_0$ and $\epsilon_n$ for $n>0$, one generically has $\epsilon_n^2\ll 1$, and the SM widths are rather suppressed. For example, the decay width of KK modes with
mass $m_n\ll m_Z$  
to a pair of SM leptons $\ell\bar{\ell}$ is~\cite{Batell:2009yf,McDonald:2010fe}
\bea
\Gamma(X_n\rightarrow \ell\bar{\ell})\ \simeq\ \frac{1}{3}\ c_W^2 \alpha\ \epsilon_n^2 m_n .\label{VectorsToSM}
\eea
Similarly, the width to hadrons is
\bea
\Gamma(X_n\rightarrow \mathrm{hadrons})
~=~\Gamma(X_n\rightarrow \mu^+\mu^-)\,\times R(s=m_n^2),
\eea
where $R(s)$ is the usual hadronic $R$ parameter, 
\mbox{$R=\sigma(e^+e^-\rightarrow
  \mathrm{hadrons})/\sigma(e^+e^-\rightarrow \mu^+\mu^-)$} \cite{Amsler:2008zzb,Ezhela:2003pp}. 
The lowest data point in the hadronic cross section data 
set is at $\sqrt{s}=0.36$~GeV, well above the pion threshold, enabling one to
use the $e^+e^-\rightarrow\pi^+\pi^-$  cross section for
the region above 
threshold~\cite{Ezhela:2003pp,Davier:2002dy,Batell:2009yf}. Note that, beyond the $\mathcal{O}(1)$ growth in the number of kinematically accessible SM states,  the total
width to the SM is roughly independent of mode number;
the KK masses grow like $m_n\sim n$, while $\epsilon_n^2\sim 1/n$, giving $\epsilon_n^2m_n\sim \mathrm{constant}$.
Provided 
$\epsilon_0 \gtrsim 10^{-4}$ the SM decays are relatively prompt on collider 
timescales ($c\tau < 1\,mm$) for $R^{-1} \sim$~GeV~\cite{Reece:2009un}.

The couplings of KK gravitons to the UV-localized SM and DM are highly
suppressed. Similar to RS2 models~\cite{Randall:1999vf}, the KK gravitons induce a
sub-dominant correction to the Newtonian potential, but
play no direct 
role in, e.g., collider physics. The couplings between KK gravitons and
KK vectors, on the other hand, are not suppressed, and the KK gravitons
play an important role in the phenomenology of the KK vectors. In
particular, KK
vectors may decay within the hidden sector via the creation of a KK
graviton and a 
lighter vector mode, $\smash{X_n\rightarrow X_m h_a}$. These
decays are kinematically allowed for sets of KK numbers satisfying
$n>m+a$. Therefore, all vector modes
with $n\ge 2$ can decay via KK graviton production. 

The total
two-body decay width of
the $n$th mode due to graviton production is
found by summing over all allowed final states,
\bea
\Gamma_n \equiv \sum_{a,m}\Gamma(X_n\rightarrow h_a X_m).\label{total_2_body_vector_decay}
\eea
For $n\gg1$ a simple parametrization of this two-body hidden decay width is
possible~\cite{McDonald:2010fe},
\bea
\Gamma_n\ \simeq\ \frac{k^3}{M_*^3}\ \frac{n^3}{8\pi}\ m_n \quad\quad\mathrm{for}\quad\quad n\gg 1 .\label{width_parama}
\eea 
An $n$-dependent parametrization
for the hidden decays of lighter modes ($n=2,3,4,5$) was
also given in Ref.~\cite{McDonald:2010fe}. Comparing the widths for SM decays to the hidden sector widths $\Gamma_n$, one finds that, when available, decays
to the hidden sector are dominant. Indeed, the branching fraction to
the SM is already small for the mode $n=2$ and is even smaller
for larger $n$. Thus for $n\ge 2$ the KK vectors
decay predominantly within the hidden sector, while the $n=0$ and
$n=1$ modes, which cannot decay via KK graviton production, decay to
the SM. Also note that the $n^4$-dependence of \eqref{width_parama} is
readily understood: The decay vertex involving gravitons has mass-dimension negative-one and thus the width requires a factor of $m_n^2$. In addition, the dominant decays are those with $n\sim m+a$ (see below), so the sum over final states in \eqref{total_2_body_vector_decay} gives a factor of $\sim n$. Combining these factors with the standard dependence on the mass of the decaying particle gives the $n^4$-dependence. 

It is clear from Eq.~\eqref{width_parama} that the KK modes rapidly become broad for $n\gg1$  so that the KK description breaks down unless $k/M_*$ is
hierarchically small. This marks the transition from the ``KK regime''
of well defined individual KK resonances (dual to light CFT
composites), to the ``continuum regime'' in which the states are
broad and overlapping and form an effective continuum (dual to the CFT
regime). For the latter modes, incalculable multi-body decays and
final states involving stringy-excitations can be important~\cite{Csaki:2008dt,Reece:2010xj}. Note, however, that for energies $E\ll M_{Pl}$, inclusive cross sections for KK-production calculated locally on the UV brane will be reliable as the local 4D cutoff is $\sim M_{Pl}$.

The use of an explicit IR localized Higgs$'$ to break $U(1)_\x$ can modify the decay properties of the KK modes by opening up new  hidden-sector decays like $X_n\rightarrow h'+X_m$. These serve to increase the overall width of the higher KK modes but do not change the fact that the dominant decay channels are within the hidden sector. On the other hand,
the $n=1$ mode, which has no kinematically available decay channels involving KK gravitons, can develop a hidden-width if the Higgs$'$ is light. When available, this hidden width typically dominates the SM width as the latter proceeds through the gauge kinetic-mixing and is suppressed by $\epsilon^2_{n=1}\ll 1$. Provided $m_0+m_{h'}<m_1$, the decay $X_1\rightarrow X_0+h'$ is therefore available and will typically 
dominate  SM decays like $X_1\rightarrow \ell\bar{\ell}$. 

Ultimately the Higgs$'$ decays to the SM to produce a collection of light SM fields. However, the kinematics and multiplicities of the SM fields depend on the value of $m_{h'}$. If the Higgs$'$ is long-lived, the expected signal can include a displaced vertex or even missing energy. This
behavior is similar to that of a light hidden-Higgs in $U(1)$ models~\cite{Batell:2009yf}.   Specifically, for $m_{h'}>2m_0$ the Higgs$'$ 
 decays (promptly) to a pair of zero mode vectors, $h'\rightarrow 2X_0$,
which in turn decay to light SM fields. Alternatively if $2m_0>m_{h'}>m_0$, the Higgs$'$ 
decays to an off-shell vector, $h'\rightarrow X_0+X_0^*$, again
giving an observable final state of four light SM fields as both
vectors produce SM fields. However, as
$m_{h'}$ approaches the lower part of this range (from above) the decay lengths
increase and displaced vertices are expected. For even
smaller masses, $m_{h'}<m_0$, the dominant Higgs$'$  decay is into
two off-shell zero modes that produce either four light SM fields via
a tree diagram, or
two SM fields via a loop diagram. In this case the decay
lengths can exceed one meter and result in a missing energy signal (for typical parameters and small
values of $m_{h'}$), similar to standard 4D results~\cite{Batell:2009yf,Gopalakrishna:2008dv}.  

The KK spectrum also contains a single gravi-scalar fluctuation (the radion)  that behaves much like a hidden Higgs. When kinematically available, the radion induces hidden-sector decays like $X_n\rightarrow r+X_m$. As with the Higgs$'$, radion decays will typically dominate the decay width of $X_1$ when available. The value of the radion mass depends on the stabilization
mechanism for the length of the extra dimension. For Goldberger-Wise
stabilization in
the calculable regime, the radion is typically on the order of the KK
scale~\cite{Csaki:2000zn}. Heavier masses are possible in the case of a larger backreaction. In this work we follow Ref.~\cite{McDonald:2010fe} and focus on the case when radion decays are not kinematically available for $X_1$ and $X_0$.

For a given KK level $n$, the partial
width for graviton decays,
$\Gamma(X_n\rightarrow h_a X_m)$, varies substantially as one varies
the daughter KK numbers $a$ and $m$~\cite{McDonald:2010fe}. Decays to
daughters $h_a$ and $X_m$ with quantum numbers satisfying $a+m\sim n$ are dominant, so
 KK number is approximately conserved. If kinematically
permitted the daughter vector will, in turn, decay into lighter
gravitons and vectors, and the daughter graviton will decay back to
two lighter vectors. Thus the creation of a heavier KK vector produces a
cascade decay down the KK towers in the hidden sector until one ends up
with a collection of light KK vectors, which in turn
decay to light SM fields. This feature in the 5D theory is dual to the showering of CFT  modes in the dual 4D theory; a heavy CFT state showers until the decay products have energies and masses on the order of the mass gap, at which point the only available decay channels are to the SM. 

The bounds on a tower of hidden KK vectors were detailed in Ref.~\cite{McDonald:2010fe}. Provided that $k\gg \mathcal{O}(\mathrm{TeV})$, the bounds on the tower of KK vectors when 10~MeV~$\lesssim m_0 \lesssim$~GeV are 
essentially the same as those on a single hidden vector with the
same mass and mixing as the zero mode~\cite{McDonald:2010fe}. This is
because important low-energy constraints, like leptonic magnetic
moments and beam dump experiments, typically weaken as the mass of the
vector increases. Also, the  kinetic mixing strength between the zero
mode and the SM is larger than that of the higher modes,
$(\epsilon_n/\epsilon_0)^2\simeq[n\times \log (kR)]^{-1}\ll1$. For
values of $m_0\simeq\mathcal{O}(10-100)$~MeV, the most important
constraints come from the lepton magnetic moments and Upsilon
decays and require
$\epsilon_0\lesssim 10^{-3}$~\cite{Fayet:2006sp,Pospelov:2008zw,Bjorken:2009mm,Hook:2010tw}. For lighter values of the zero-mode mass
the constraints on $\epsilon_0$ from beam dumps and supernova cooling are more
severe (the latter bounds were recently updated for a single hidden vector~\cite{Dent:2012mx}). The bounds on a full KK tower of hidden vectors have not been studied in detail for $m_0\lesssim 10$~MeV. However, we expect that these bounds will be at least as severe as those for a single hidden photon, which are at the level of $\epsilon\lesssim10^{-10}$. This provides a rough guide for the bound on $\epsilon_0$ for $m_0\simeq\mathcal{O}(1-10)$~MeV, though the reader should bear in mind that the constraints for the full tower may be more severe.

The mixing between the SM $Z$ and the KK vectors induces
a small shift in $m_Z$. However, this is within experimental
uncertainties once $\epsilon_0$ satisfies the more severe low-energy
constraints~\cite{McDonald:2010fe}. The gauge kinetic mixing also permits
exotic $Z$ decays, with typical final states consisting of a large number
of soft SM fields. These are created when the $Z$ mixes with a KK mode with
mass $m_n\simeq m_Z$, which then showers within the hidden sector to produce
a collection of light KK vectors. The latter then decay to the
SM. Specific signals of this type were not explicitly searched for at
LEP, so no direct bound can be given~\cite{McDonald:2010fe}; note that the bound on
the invisible
$Z$ width is not applicable as the final state consists
of SM fields. Also note that the inclusion of $U(1)_{\x}$ charged DM,
which was not
considered in Ref.~\cite{McDonald:2010fe}, can result in more
stringent bounds. In particular, if the DM is a Dirac
fermion, direct detection searches require $\epsilon_0\lesssim
10^{-6}(m_0/\mathrm{GeV})^2$~\cite{Angle:2007uj,Dedes:2009bk}. This bound does not
apply if the DM scatters inelastically, i.e.~if the Dirac fermion is
split into two 
near degenerate Majorana fermions. 

We focus on lighter IR scales of $\Lambda_\ir\lesssim10$~GeV in this work, but the region of parameter space with $\Lambda_\ir>m_Z$ is potentially interesting with regards to the LHC. Although the bounds for larger values of $\Lambda_\ir\gtrsim10$~GeV have not been studied in detail, it is clear that the hidden-sector $Z$-decays will vanish for $\Lambda_{\ir}\gtrsim m_Z$, and that other important low-energy constraints, like the muon magnetic moment, will be less important than, e.g., electroweak precision tests~\cite{Langacker:2008yv,Chang:2006fp}. The bounds on the kinetic mixing with a single ``shadow vector'' are on the order of $\epsilon\lesssim 10^{-3}$ for $M\sim10^2$~GeV, and decrease with increasing mass~\cite{Chang:2006fp}.  For $\Lambda_\ir\gtrsim10^2$~GeV the KK tower would act like a tower of ``shadow vectors'' and the bounds are expected to be at least as severe as those in Ref.~\cite{Chang:2006fp}. 

Before turning our attention to DM annihilations, we note that one of the key
differences between models that stabilize the hidden $U(1)$ symmetry breaking scale with
warping/compositeness, compared to other approaches like SUSY~\cite{Abel:2008ai,ArkaniHamed:2008qp,Baumgart:2009tn}, is the nature of the expected
low-energy phenomenology. The multiple light KK vectors lead to a rich
phenomenology~\cite{McDonald:2010fe} which could manifest in on-going
low-energy experiments~\cite{Bjorken:2009mm}. Of particular interest
is the fact that resonant behavior from multiple light KK modes could
appear in low-energy experiments and that, in some cases, signals could
already be present in existing data sets. For example, the warped
model predicts anomalous production of six- and eight-lepton final states at low energy
colliders like the $B$-factories~\cite{McDonald:2010fe}, due to
processes like $e^+e^-\rightarrow h_1 X_0\rightarrow 2\ell \, 4\ell^\prime$. Production
occurs through the $s$-channel and is only suppressed as $\epsilon^2$. The
signal can even be resonantly enhanced if 
a given experiment operates with a
center-of-mass energy near one of the light-KK masses,
$\sqrt{s}\simeq m_n$ for small $n$~\cite{McDonald:2010fe}. The specific six- and eight-lepton final
states predicted by the warped model have not been explicitly
searched for to date, and  an analyses of $B$-factory data sets for multi-lepton signals could provide more stringent bounds on (or discover evidence for) the current model. Note that the BABAR collaboration~\cite{:2009pw} has
performed an
analysis of four-lepton final states, $e^+e^-\rightarrow 2\ell \, 
2\ell^\prime$, which occur in models with a non-Abelian hidden sector (via $e^+e^-\rightarrow W' W'$ when the vectors decay to the SM; see e.g.~\cite{Carloni:2011kk}). A rather strong bound of $\sigma(e^+e^-\rightarrow 2\ell 2\ell')<(25-60)~ab$ for vector masses in the range $[0.24,5.3]$~GeV was obtained.
In the present model, anomalous four-lepton production occurs through the $t/u-$channel and is suppressed as $\epsilon^4$, so the bound is not severe. 

 Let us also note that recently reported results (and ongoing experiments) by the A1 Collaboration at Mainz~\cite{Merkel:2011ze}, KLOE~\cite{Archilli:2011nh}, and the APEX Collaboration at JLAB~\cite{Abrahamyan:2011gv}, are probing the interesting range of parameter space for a lightest KK vector with $m_0\simeq\mathcal{O}(100)$~MeV and $\epsilon_0\lesssim10^{-3}$. Future results from these groups will further explore this parameter space and offer the potential of discovering the lightest KK modes. The potential for
interesting low-energy 
phenomenology, at both fixed target experiments and low-energy colliders, is a core feature of warped models with a light
hidden sector.
\section{Dark Matter Annihilations\label{sec:DAannihilations}}
With the above information we can now determine the annihilation properties of
the UV-localized DM. The annihilation spectrum depends on the
coupling of the DM to the individual KK modes. As mentioned already, we take a $U(1)_{\x}$ charge of $Q_\psi=+1$ for the
DM, though
the specific value is not important as one can always rescale the bulk
gauge coupling.  The action for $\psi$ is
\bea
S\supset\int d^5x\sqrt{-g} \left\{\frac{i}{2}\bar{\psi}\Gamma^\mu D_\mu \psi - M_{\dm}\bar{\psi}\psi+ H.c.\right\}\delta(z-k^{-1}),
\eea
where $\Gamma^\mu$ are the curved space Dirac matrices. Integrating over the extra dimension gives
\bea
S\supset\int d^4x \left\{i\bar{\psi}\gamma^\mu \partial_\mu \psi
  +\sum_n g_{n}\bar{\psi}\gamma^\mu \psi X_\mu^{n}- M_{\dm}\bar{\psi}\psi\right\},
\eea  
where $\gamma^\mu$ are the usual 4D Dirac matrices. The effective 4D coupling between $\psi$ and the $n$th KK mode ($g_n$) is defined in terms of the
5D gauge coupling ($g_5$):
\bea
g_{0}&=&g_5\ f_0(k^{-1}) \simeq - \frac{g_5\ \sqrt{k}}{\sqrt{\log(2k/m_0)-\gamma}}\ ,\nonumber\\
g_{n}&=&g_5\ f_n(k^{-1}) \simeq -\frac{g_5\ \sqrt{k}}{[\log(2k/m_n)-\gamma]}\,(n+1/4)^{-1/2}\ ,~~~{n \geq 1}.
\label{DM_coupling_n}
\eea
As with the kinetic mixing parameters, we can approximately relate the effective 
coupling constant for the light KK modes to that of the zero mode: 
\bea
g_{n}\simeq
\frac{1}{\sqrt{n}}\frac{g_{0}}{\sqrt{\log(2k/m_n)-\gamma}}\ .\label{brane_coupling_rel_single}
\eea
Thus the coupling of DM to the zero mode is dominant. As an example, with $R^{-1}\sim$~GeV, one finds $g_n/g_0\sim 1/(6\sqrt{n})$. 

Note that to obtain Majorana DM an additional ingredient must be added to generate the small mass-splitting. If there is a bulk Higgs$'$ that is localized towards the IR brane and has $U(1)_\x$-charge $2Q_\psi$, the DM will couple to the UV value of the Higgs$'$ field via a bilinear Majorana term. After symmetry breaking this can generate a small ($\ll \Lambda_{\ir}$) Majorana mass for $\psi$ and split the Dirac fermion into two Majorana fermions. Alternatively, if symmetry breaking occurs purely on the IR brane, a bulk mediator must be added to communicate with $\psi$ and generate the mass split. We will not consider this matter further here as these model building details are not important for our main points.

\begin{figure}[ttt]
\begin{center}
\includegraphics[width = 0.3\textwidth]{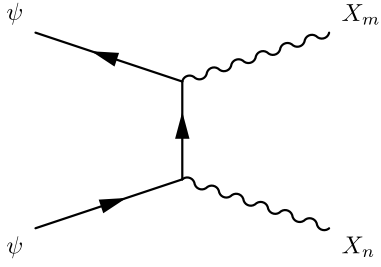}
\end{center}
\caption{Dark matter annihilates into two hidden Kaluza-Klein
  vectors.}
\label{DM_ann_KK}
\end{figure}
There are two main ways in which the DM can annihilate into the hidden
sector. The $t$-channel involves the
DM annihilating into two
KK vectors, $\psi\bar{\psi}\rightarrow X_nX_m$, as shown in Fig.~\ref{DM_ann_KK}. The  cross
 section for zero mode production is of the standard form:
\bea
\sigma( \psi\bar{\psi}\rightarrow X_0X_0)\times v&\simeq& \frac{1 }{16\pi
  }\frac{g_{0}^4}{
  M_{\dm}^2}.\label{dm_ann_2X0}
\eea
where $v$ is the magnitude of the relative velocity between the
initial state particles.  We
can use Eq.~(\ref{brane_coupling_rel_single}) to approximately
relate the cross sections for
different final states:
\bea
\frac{\sigma (\psi \bar{\psi}\rightarrow X_n X_0)}{\sigma (\psi
  \bar{\psi}\rightarrow X_0 X_0)}&\simeq & \frac{2g^2_n}{g^2_0}\sim
\frac{1}{20 n}\ ,\quad n\ne0\ ,
\label{BR00-0n}
\eea
and
\bea
\frac{\sigma (\psi \bar{\psi}\rightarrow X_n X_m)}{\sigma (\psi
  \bar{\psi}\rightarrow X_0 X_0)}&\simeq &
\frac{2g^2_mg_n^2}{g_0^4}\sim \frac{1}{700}\frac{1}{ nm}\ ,\quad n \ne
m>0\ ,
\label{BR00-mn}
\eea
where the numerical values hold for $R^{-1}\sim$~GeV, but similar results obtain for smaller $R^{-1}$. Observe that the annihilation into two zero-modes
 is dominant. Also note that we
 have neglected the final-state vector masses and the corresponding reduction in
 available phase space
 for annihilation into modes with $m,n\ne0$. 

The total $t$-channel annihilation cross section is found by summing over all kinematically allowed final-state KK vectors. We can estimate this inclusive cross section as
follows. The cross section for
annihilation to the $m$th and $n$th gauge boson is approximately
\bea
\sigma_{mn}\equiv \sigma(\psi\ \bar{\psi}\longrightarrow X_m\, X_{n})\sim\frac{1}{16\pi}\frac{g_5^4}{vM_{\dm}^2}[f_m(k^{-1})]^2[f_n(k^{-1})]^2,
\eea
and the inclusive cross section is obtained by summing over all kinematically
allowed final states,
$\sigma_{\tiny{total}}=\sum_{m,n}\sigma_{mn}$. This sum can be split
into two parts, $\sigma_{\tiny{total}} =\sigma_<+\sigma_>$, where
$\sigma_<$ is
the sum over the small-$n$ modes, which are narrow and well-defined KK
modes, and $\sigma_>$ is the sum over the large-$n$
modes, which are broad, overlapping
resonances displaying RS2-like continuum behavior. For these latter
modes we may convert the sums to integrals and take the RS2 limit:
\bea
\sigma_>\sim \frac{1}{16\pi}\frac{g_5^4 k^2}{vM_{\dm}^2}\int_{m_5}^{M_{\dm}}dm_m\int_{m_5}^{M_{\dm}}dm_n\frac{1}{m_m[\log(k/m_m)]^2}\frac{1}{m_n[\log(k/m_n)]^2},
\eea
where we have used the UV value of the bulk vector wavefunction in
RS2 and cut the integrals off in the UV at
$M_{\dm}$ and in the IR at the point where the KK modes become broad and overlap; i.e., at the point where the KK modes display RS2-like behavior. We take this point to be at the $n=5$ mode but the result is not particularly sensitive to this choice.\footnote{More generally, the bulk 5D vector becomes RS2-like for UV injection energies $E\gg (M_*/k) R^{-1}$.}  Performing the integrals and writing the result in terms of the zero-mode coupling gives
\bea
\sigma_>&\sim&\frac{1}{16\pi}\frac{g_0^4}{vM_{\dm}^2}\left[\frac{\log(k/m_0)\log(M_{\dm}/m_5)}{\log(k/M_{\dm})\log(k/m_5)}\right]^2.\label{heavy_modes_approx_tu}
\eea
The log factors suppress $\sigma_>$ relative to $\sigma_{00}$ provided $k/M_{\dm}\gg M_{\dm}R$. For example, one obtains $\sigma_>\sim \sigma_{00}/50$ for $M_{\dm}\sim$~TeV, $R^{-1}\sim$~GeV (taking
$m_5/m_0\sim 10$, which always holds). With Eq.~(\ref{heavy_modes_approx_tu}) we deduce 
that the inclusive cross section in the
$t/u$-channel is dominated by the annihilation to the lighter, narrow modes (chosen as $n<5$ here), which is itself dominated by zero-mode production. Thus
$\sigma_{\tiny{total}}\sim \sigma_<\sim \sigma_{00}$, as expected on
the basis of the coupling relations (\ref{brane_coupling_rel_single}),
and  the $t$-channel is dominated by zero-mode
production. Note that one can also estimate the cross section for production of modes with masses
between some energy scale $E\gg R^{-1}$ and $M_{\dm}$ by replacing $m_5\rightarrow E$
in Eq.~(\ref{heavy_modes_approx_tu}). For example, one finds that
the cross section for production of modes with masses in the range $500~\mathrm{GeV}\lesssim
m_n\lesssim M_{\dm}\sim$~TeV is $\sigma\sim 10^{-3}\times \sigma_{00}$ for $R^{-1}\sim $~GeV.

In addition to the creation of two KK vectors, the DM can annihilate into a single \emph{on-shell} KK vector that escapes from the brane into the bulk, $\psi\bar{\psi}\rightarrow X_n$ with $m_n\simeq 2M_{\dm}$. This inverse decay dominates the other $s$-channel hidden-sector production processes, which proceed through an off-shell vector. Note that the KK spectrum is discrete and naively one would not be guaranteed that it contains a mode with $m_n\simeq 2M_{\dm}$. However, for $n\gg1$ the KK modes form an effective continuum as the resonances are broad and overlapping; this is equivalent to the statement that the dual 4D theory is conformal at energies $E\gg \Lambda_{\ir}\sim R^{-1}$. Thus, in the region of parameter space defined by $M_{\dm}R\gg1$, the inverse decay is always kinematically available. 

\begin{figure}[t]
\begin{center}
        \includegraphics[width = 0.3\textwidth]{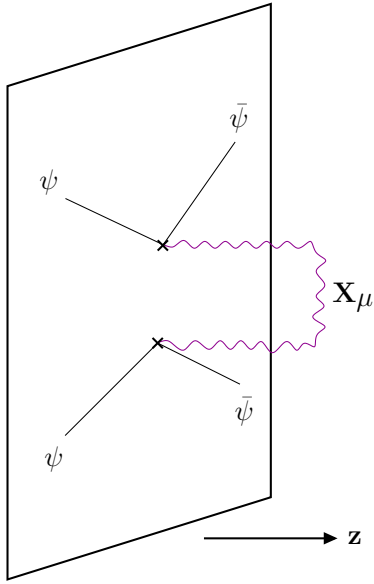}
\end{center}
\caption{Dark matter localized on the UV brane (represented by the
  plane) annihilates into a vector $X_\mu$ that
  propagates into the bulk before returning to the UV brane. The inclusive cross
  section for bulk field production in the $s$-channel
  can be calculated 
  by applying
  a unitarity cut to this diagram.}
\label{Fig:dm_s}
\end{figure}

Note that, in practise, the KK vector created via $\psi\bar{\psi}\rightarrow X_n$ will rapidly shower in the hidden sector, creating lighter KK vectors and gravitons (and possibly stringy-resonances), producing a complicated multi-body final state. The precise nature of this final state cannot be determined but one can reliably compute the inclusive cross section for $s$-channel hidden-sector production. This is achieved by performing a unitarity cut on the bulk-vector propagator in the process shown in Fig.~\ref{Fig:dm_s}. For
$M_{\dm} R\gg 1$, the RS1 bulk vector propagator is well
approximated by the bulk RS2 propagator, and we therefore use the latter for the calculation.\footnote{This is true provided $k/M_*$ is not
hierarchically small. More generally, the UV-to-UV RS1 bulk vector propagator
approaches the RS2 result for $M_{\dm} \gg (M_*/k)R^{-1}$. For more discussion of the
propagator matching see Ref.~\cite{McDonald:2010fe}.}   For completeness, the coefficient of the gauge-invariant transverse 
part of the RS2 UV-to-UV propagator is~\cite{Dubovsky:2000am,Friedland:2009iy}
\bea
\Delta_p^{RS2}(k^{-1},k^{-1}) &=& \frac{H_1^{(1)}(p/k)}{pH_0^{(1)}(p/k)}
\label{rs2prop}
\\
&\simeq& 
\frac{k}{p^2[\log(2k/p)-\gamma]}\times
\left\{
\begin{array}{ccc}
\left(1
-i\frac{\pi}{2}/[\log(2k/p)-\gamma]\right)&;&p^2>0\\
\phantom{i}&\\
- 1
&;&p^2<0 \, ,
\end{array}
\right.
\nonumber
\eea
where $H_n^{(1)} = (J_n+iY_n)$.
Note that the propagator contains an imaginary part for $p^2>0$. This
results from imposing outgoing-wave boundary conditions at 
$z\to \infty$~\cite{Dubovsky:2000am,Giddings:2000mu} and accounts for the production of a real on-shell vector that escapes into the bulk. 

Performing the unitarity cut
gives
\bea
\sigma_s (\psi \bar{\psi}\rightarrow
Hidden)\times v&\simeq&\frac{3\pi}{8
  M_{\dm}^2}\frac{g_{5}^2k}{[\log(k/M_{\dm})-\gamma]^2}\ .\label{inclusive_hidden_prod_one}
\eea
Note that the dependence on $k$ in
Eq.~(\ref{inclusive_hidden_prod_one}) is only mild
(logarithmic) and there is
no dependence on the IR scale $R$. In general the cross section possesses the latter
property whenever
$M_{\dm} \gg (M_*/k)R^{-1}$. This feature is readily understood: The
bulk vector propagator does not probe the deep IR for
production energies $E\gg (M_*/k)R^{-1}$, as it becomes highly oscillatory for $z>E^{-1}$ and essentially cuts off at $z\simeq E^{-1}$. Therefore $s$-channel hidden-sector production is insensitive to the details of the
geometry in the deep IR, as reflected in
Eq.~(\ref{inclusive_hidden_prod_one}). Note also that, although the KK theory becomes strongly coupled in the IR region towards
$z=R$ (where the local 4D cutoff is $\sim (M_*/k)R^{-1}\ll M_*$), one can still trust the
weakly-coupled bulk 5D theory for processes initiated locally on the UV brane with injection energies $E\gg (M_*/k)R^{-1}$. Such processes are controlled by the local UV cutoff, which is of order $M_{Pl}$, and as long as $E\lesssim M_{Pl}$ the weakly-coupled description remains 
reliable.

\begin{figure}[ttt]
\begin{center}
        \includegraphics[width = 0.4\textwidth]{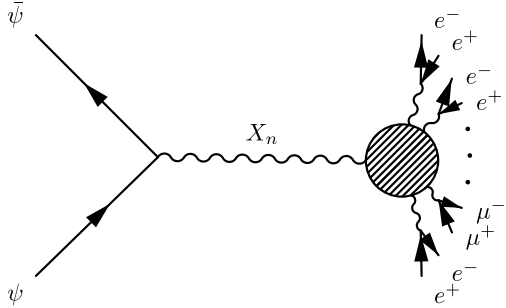}
\end{center}
\caption{Dark matter annihilation into light standard model fermions
  through the $s$-channel. The dark matter annihilates into heavy hidden
  KK modes which cascade in the hidden sector, producing
a collection of  light hidden vectors (collectively represented  by
  the blob), which in turn decay to a large number of light SM fields.}\label{DM_ann_x_SM}
\end{figure}

Although both the $s$- and $t$-channels are dominated by hidden-sector production, the specific final states are quite different for the two channels. The hidden KK vectors are not stable and ultimately decay to produce SM fields. The $t$-channel predominantly creates boosted zero-modes with energy $\sim M_{\dm}$, which decay to the SM through the gauge kinetic-mixing with hypercharge. This produces a spectrum of hard, light SM fields. $s$-channel hidden-sector
 production instead
creates heavy KK modes ($m_n\sim M_{\dm}$) that lie within the RS2-like
 part of the spectrum; i.e.~broad, overlapping resonances
 displaying continuum-like behavior. These modes rapidly shower
 within the hidden sector, producing lighter graviton and vector
 modes. The showering continues until the energy of the hidden modes
 is on the order of the IR scale, at which point
 the final state consists of a large collection ($\sim RM_{\dm}$) of soft, light KK modes.\footnote{The light states can also include the radion, depending on the value of $m_rR$.} The only decay channels available to these light
 modes are to the SM, so the final state consists of a large
 multiplicity ($\sim RM_{\dm}$) of light SM fields with energies of order $R^{-1}$ (see
 Fig.~\ref{DM_ann_x_SM}).

In addition to the annihilations into the hidden sector the DM can also annihilate directly into the SM. However, such processes necessitate a kinetic-mixing insertion between a KK vector and SM hypercharge. Given that $\epsilon_n\ll1\ \forall n$, these direct annihilations are highly suppressed relative to hidden-sector production. Thus, the dominant annihilation channels are the hidden-sector ones described above.

Having detailed the annihilation properties of the DM we can proceed to consider the early universe cosmology. We turn to this matter in the following section. Before concluding this section we note the relationship between the inclusive
$s$-channel cross section,
Eq.~(\ref{inclusive_hidden_prod_one}), and the dominant $t$-channel process $\psi\bar{\psi}\rightarrow 2X_0$:
\bea
\frac{\sigma_s(\psi\bar{\psi}\rightarrow Hidden)}{\sigma
(\psi\bar{\psi}\rightarrow X_0
X_0)}&\simeq& \frac{6\pi^2}{g_5^2k}\frac{[\log(2k/m_0)-\gamma]^2}{[\log(k/M_\dm)-\gamma]^2}\ \sim\ \mathcal{O}\left(\frac{1}{g_5^2k}\right)\ .\label{relating_s_t_channel}
\eea
This suppression of the $t$-channel relative to the $s$-channel has a clear interpretation in the dual 4D theory: The AdS/CFT dictionary gives $(g_5^2k)^{-1}\sim N$, so this ratio is sensitive to the number of colors in the large-$N$ CFT (see Section~\ref{ads_cft}).

\section{Early-Universe Cosmology\label{sec:cos}}
\begin{figure}[t]
\begin{center}
        \includegraphics[width = 0.4\textwidth]{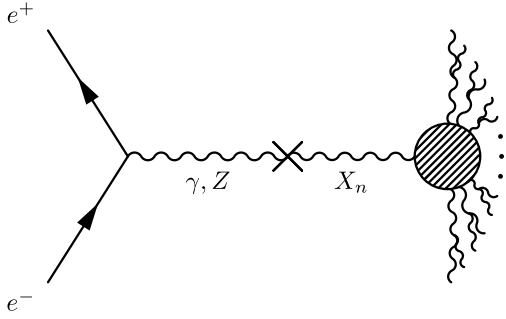}
\end{center}
\caption{Standard Model fields (like electrons) can annihilate into the hidden sector through the $s$-channel. This proceeds through the mixing between hypercharge and the hidden KK-vectors/CFT-states, and serves to keep the SM and the CFT in thermal contact.}
\label{Fig:2e_s}
\end{figure}

We now discuss the early-universe cosmology for our model. In addition to the SM fields, the thermal plasma in the early universe can contain the $U(1)'$-charged DM ($\psi$) and the hidden-sector CFT/KK states.  The SM keeps thermal contact with the hidden-sector predominantly 
via $s$-channel processes of the type shown in Fig.~\ref{Fig:2e_s}. Both sectors are in
thermal equilibrium when the dark matter freezes out (assuming that the reheating temperature is $\smash{T_R > M_\dm}$) at a temperature $T_f$  if
\bea
n_\sm(T_f)  \, \langle \sigma v \rangle \, > \, H(T_f) \, ,
\label{ThermalContactCondition}
\eea
where $\smash{n_\sm (T) \approx \frac{g_\sm}{\pi^2} T^3}$ is the particle number-density in the standard model sector with $g_\sm$ relativistic degrees of freedom.\footnote{For temperatures between 300 MeV and 1 MeV, $g_\sm \sim 10$, and for temperatures above 300 MeV, $g_\sm \sim 100$.}  If the dark matter is a cold relic, it freezes out at a temperature $T_f$ about a factor of $10$ below its mass. For example, for a standard WIMP with a weak-scale mass one finds $\smash{T_f \approx M_\dm / 20}$. The Hubble rate is $H= \smash{\sqrt{\rho/3}/M_{Pl}}$, where the total energy-density at temperatures $T$ above the IR scale $\Lambda_\ir$ is given by
\bea
\rho(T) \, = \, \frac{\pi^2}{30} g_\sm  \, T^4 + \rho_\cft(T)\, .
\eea
The contribution from the strongly-coupled CFT can be expressed in terms of the central charge of the CFT as $\rho_\cft(T)=\frac{3\pi^2}{2} c\, T^4$~\cite{Henningson:1998gx,Gubser:1999vj}. With this we write the total energy-density as 
\bea
\rho(T) \, = \, \frac{\pi^2}{30} g_*  \, T^4 \, ,
\eea
where the effective number of relativistic degrees of freedom is 
\bea
g_* \, \equiv g_\hid+\,g_\sm\, ,
\eea
and the hidden sector contribution is 
\bea
g_\hid\simeq   180 \, \pi^2 \left(\frac{M_*}{k}\right)^3.
\label{energydensity}
\eea
Note that the number of colors in the CFT is $N^2\sim (M_*/k)^3$ so we have $g_\hid\sim N^2$, as expected for the unconfined phase of a large-$N$ theory.

Ignoring the tiny coupling of the SM and the CFT due to gravity, all processes that couple the two sectors must proceed through a kinetic mixing insertion. The dominant such process coupling electrons to the CFT is shown in Figure~\ref{Fig:2e_s}. For energies $\sqrt{s}\gg\Lambda_{\ir}$, the inclusive cross section is found by following the procedure of Section~\ref{sec:DAannihilations} and performing a unitarity cut on a bulk RS2 propagator, giving
\bea
\sigma(e^+e^-\rightarrow CFT)\ \simeq\ \frac{\pi\alpha}{2c_W^2}\left(\frac{k}{M_*}\right)\frac{ \epsilon_*^2}{[\log(2k/\sqrt{s})-\gamma]^2}\frac{Y_e^2}{s},
\eea
where $Y_e$ is the electron hypercharge and we have used\footnote{For smaller (but off-resonant) values of $\sqrt{s}\sim m_Z$ the cross section will depend on the $Z$-coupling rather than the hypercharge coupling, while for $\sqrt{s}\ll m_Z$ the relevant coupling is the electric charge~\cite{McDonald:2010iq}.} $\sqrt{s}\gg m_Z$. Note that this result is insensitive to the details of the geometry in the IR (equivalently, it does not depend on the confinement scale). Summing over the relevant SM fields, the cross section for annihilation of SM particles to the CFT is estimated as 
\bea
\sigma(SM\rightarrow CFT)\ \simeq\ \frac{\pi\alpha}{2c_W^2}\ \frac{\epsilon_0^2}{s}\ \frac{\log(2k/m_0)-\gamma}{[\log(2k/\sqrt{s})-\gamma]^2}\times \sum_iY_i^2.\label{sm_cft_cross_section}
\eea
Note that the dependence of \eqref{sm_cft_cross_section} on the confinement scale (through $m_0$) is artificial and results from expressing the cross section in terms of the kinetic-mixing parameter relevant for low-energy experiments ($\epsilon_0$).

 Combining these results, the SM is in equilibrium with the CFT at the (thus far assumed) DM freeze-out temperature, $T_f\sim M_{\dm}/20$, provided the kinetic-mixing parameter satisfies\footnote{We use $\sum_iY_i^2=10.25$ and note that the log-factor in \eqref{sm_cft_cross_section} gives $\sim1/25$ for $M_\dm\sim \mathcal{O}(1-10)\text{ TeV}$, $m_0\in[1~\mathrm{MeV},1~\mathrm{GeV}]$, and $k\sim M_{Pl}$. The slow logarithmic factors are easily restored.} 
\bea
\epsilon_0>7\times 10^{-9}\times \left(\frac{M_{\dm}}{500~\mathrm{GeV}}\right)^{1/2}\times \left(\frac{100}{g_*}\right)^{1/4}.\label{sm_cft_equil}
\eea
Up to slow logarithmic corrections, this result holds for arbitrary
$M_\dm/\Lambda_{\ir}\gg 1$ with $M_\dm\gg m_Z$. Noting the dependence on the number of colors in the CFT, through the dependence of $g_*$ on $M_*/k$, we write \eqref{sm_cft_equil} as
\bea
\epsilon_0\ >\ \kappa_{\epsilon}\times 10^{-10}\times \left(\frac{M_{\dm}}{500~\mathrm{GeV}}\right)^{1/2},\label{sm_cft_equilibrium}
\eea
where 
\bea
\kappa_{\epsilon}\ \simeq\ \{ 0.8,\ 6,\ 34\}\quad\mathrm{for}\quad \left(\frac{k}{M_*}\right)\ =\ \{10^{-2},\ 10^{-1},\ 1\}.
\eea

For a large part of the parameter space of interest in low-energy searches for hidden photons, namely $m_0\in [10~\mathrm{MeV}, 1~\mathrm{GeV}]$, typical bounds on the KK mixing-parameter are at the level of $\epsilon_0\lesssim 10^{-3}$~\cite{McDonald:2010fe}. Thus, for this mass range, Eq.~\eqref{sm_cft_equilibrium} shows that ample parameter space exists for which the CFT is in equilibrium with the SM and the hidden-vectors are consistent with constraints. For lighter values of $\smash{m_0\sim 1-10}$~MeV, the constraints on $\epsilon_0$ due to beam dump experiments and supernova cooling are more severe; the latter are at the level of $\epsilon<10^{-10}$ for mixing with a single hidden-photon~\cite{Dent:2012mx}. A detailed study of the bounds on a full tower of hidden vectors has not been performed for $m_0\lesssim10$~MeV. However, we expect these to be at least as severe as the bounds on a single hidden-photon. Assuming the (possibly optimistic) bound of $\epsilon_0<10^{-10}$, we see that the SM and the CFT cannot be in equilibrium for
\bea
1\lesssim (m_0/\mathrm{MeV})\lesssim10\ \quad \mathrm{and}\quad M_\dm\gtrsim~\mathrm{TeV}\quad\mathrm{when}\quad (k/M_*) \gtrsim10^{-2}.
\eea
Therefore, for  $m_0\sim $~MeV, the ``smaller''-$N$ limit of $k/M_*\rightarrow\mathcal{O}(1)$ is incompatible with CFT-SM equilibrium in the early universe. Regions of the (very) large-$N$ parameter space, $\smash{M_*/k\gtrsim 10^{3}}$, might remain compatible, subject to the precise nature of the supernova bounds on the full tower.

With the SM and the CFT in equilibrium, the CFT will, in turn, keep the DM in equilibrium, provided the gauge coupling $\smash{g_5\sqrt{k}}$ is not extremely small (as we quantify below). The relic abundance of the DM is then determined by the usual freeze-out process. We can obtain an order-of-magnitude estimate for the relic abundance by solving the freeze-out condition for the particle number density:
\bea
n_\dm (T_f) \, \sim \, \frac{H(T_f)}{\langle \sigma v \rangle} \, \propto \,  \frac{\sqrt{g_*}\,  T_f^2}{\langle \sigma v \rangle}\, , 
\eea
where the annihilation cross section of DM is given in Eq.~\eqref{inclusive_hidden_prod_one}.\footnote{Depending on the IR scale (or equivalently the mass of the gauge bosons), the annihilation cross section of non-relativistic particles can be enhanced by the Sommerfeld effect. Because the dark matter still has high velocities during freeze-out, the Sommerfeld effect typically gives
small corrections to the relevant cross section. But in extreme cases, these corrections can enhance the cross section by an order of magnitude \cite{Zavala:2009mi,Dent:2009bv}. A correspondingly lower tree-level cross section is then needed to obtain the right relic abundance.} To determine the present-day contribution $\Omega_\dm$ of DM to the critical energy density, we must evolve the number density to the present epoch. Assuming that the total comoving entropy density (i.e.~in both the CFT and SM sector) remains constant over the evolution of the universe, the ratio of the DM abundance $n_\dm$ to the entropy density $\smash{s \sim \rho/T}$ also remains constant (we discuss the conditions under which this is a good approximation below). Noting that $T_f \propto M_\dm$, up to logarithmic corrections, we thus find
\be
\Omega_\dm \, \propto \,  \frac{n_\dm}{s}\, M_\dm \, \propto \,  \frac{1}{\sqrt{g_*}  \, \langle \sigma v \rangle} \, .
\label{omegaDM}
\ee 
Notice that the explicit dependence on $M_{\dm}$ has dropped out (though there is, of course, still an implicit dependence via the annihilation cross section). This is just the well-known result that the observed mass density of DM is obtained if the annihilation cross section has a certain value, which, for a WIMP with dominant $s$-wave annihilations, is ${\langle \sigma v \rangle \approx 3 \cdot 10^{-26} \, \text{cm}^3/\text{s}}$. In our case, however, the additional degrees of freedom in the CFT (equivalently, the dual KK modes) contribute to the energy and entropy densities during freeze-out. As follows from Eq.~\eqref{omegaDM}, an annihilation cross section\footnote{We discuss the derivation of this result in more detail in Appendix~\ref{app:DM_entopy}.}
\be
\langle \sigma v \rangle \, \approx \, 3 \cdot 10^{-26} \sqrt{\frac{g_{\sm}}{g_*}} \,\,  \frac{\text{cm}^3}{\text{s}}
\label{DMcs}
\ee
is then needed to obtain the right abundance. Depending on the parameter $k/M_*$, which determines $g_*$ via Eq.~\eqref{energydensity}, this can be considerably smaller than the usual WIMP result. The dependence on $\smash{g_*}$ is easily understood: The universe expands faster during freeze-out due to the CFT, leading to a factor $\smash{\sqrt{g_*/g_{\sm}}}$ larger freeze-out abundance.  But, at later times, the CFT transfers its entropy to the SM sector, diluting the DM abundance and leading to a smaller value of $\smash{n_\dm/s}$ by factor of $\smash{g_{\sm}/g_*}$.

Taking the above results together, we thus require the coupling and mass for a thermal DM candidate to satisfy\footnote{We again use the fact that $[\log(2k/m_0)]/[\log(k/M_{\dm})]^2\approx1/25$ for the regions of interest; the dependence on the log factors is easily restored.}
\bea
\alpha_0 \simeq 1.5\times 10^{-3}\times \sqrt{\frac{g_{\sm}}{g_*}}\times \left(\frac{M_{\dm}}{500~\mathrm{GeV}}\right)^2,
\eea
where we express the result in terms of the zero-mode fine-structure constant $\alpha_0\equiv g_0^2/4\pi$. To get a feeling for the dependence on $k/M_*$ (equivalently, $N$), we rewrite this as
\bea
\frac{\alpha_0}{\alpha}\ \simeq\  \kappa_\alpha\times 10^{-2}\times \left(\frac{M_{\dm}}{500~\mathrm{GeV}}\right)^2,\label{DM_CFT_therm_condition}
\eea
where the parameter $\kappa_\alpha$ encodes the $(k/M_*)$-dependence:
\bea
\kappa_\alpha \ \simeq\ \{0.005,\ 0.1,\ 5\}\quad\mathrm{for}\quad \left(\frac{k}{M_*}\right)\ =\ \{10^{-2},\ 10^{-1},\ 1\},
\eea
and we have divided by the electromagnetic fine-structure constant for comparison. This constraint demonstrates one of the most marked effects of the hidden CFT; the large-$N$ limit (equivalently, the small $k/M_*$ limit) permits heavier DM candidates for a given fixed value of the perturbative coupling. For example, with $\alpha_0 =\alpha\, (2\alpha) $ and $k/M_*= 0.1$, we find that DM with mass $M_{\dm}\approx10\, (14)$~TeV will produce the correct abundance --- in contrast with the value $M_{\text{\tiny{WIMP}}}\approx 0.3\, (1)$~TeV for a standard WIMP with coupling $\alpha_{\text{\tiny{WIMP}}}=\alpha\, (2\alpha)$. 

Note that for $M_{\dm}/\Lambda_{\ir}\gg1$, DM freeze-out occurs at a temperature $T_f\gg \Lambda_{\ir}$. Accordingly, the CFT is in the unconfined, conformal phase during freeze-out. This phase of the 4D theory has a dual counterpart  ---  when an RS model is heated to temperatures above the IR scale, the backreaction of the thermal plasma on the geometry is strong and the IR brane is replaced by a black hole horizon~\cite{ArkaniHamed:2000ds}. The resulting geometry is known as AdS-Schwarzschild. When the temperature drops 
below the IR scale due to the expansion of the universe, a phase transition 
takes place and the black-hole horizon is replaced by an IR brane~\cite{Creminelli:2001th}. This is dual to the confining phase transition of the (broken) CFT. 

Let us discuss this phase transition in more detail.  At the critical temperature
$T_c\lesssim \Lambda_{\ir}$ (the precise value depends on the details of the model), the free energies of the confined and deconfined phases (equivalently, the AdS and AdS-Schwarzschild phases) are equal. At lower temperatures, the confined phase is thermodynamically preferred. For the large-$N$ gauge theories dual to $AdS_5$ theories, the resulting phase transition is first order (see e.g.~\cite{Creminelli:2001th,Kaplan:2006yi}) and thus proceeds via bubbles of the new phase which nucleate in the deconfined plasma. Since this process competes with the expansion of the universe, the phase transition only completes if the rate of bubble nucleation (per Hubble volume) becomes larger than the Hubble rate. If the rate of bubble nucleation is not large enough the universe remains stuck in the false vacuum.\footnote{This is similar to the graceful-exit problem in old inflation.} 

The conditions under which the phase transition can successfully complete have been studied in the literature~\cite{Creminelli:2001th,Kaplan:2006yi,Randall:2006py,Nardini:2007me,Konstandin:2010cd,Konstandin:2011dr}. The consensus is that for the case of stabilizing dynamics with a small backreaction, a somewhat stringent upper bound is typically placed on $M_*/k$ (equivalently, $N$). This bound is such that, at best, the phase transition to the RS phase only completes in the region where the gravity description is starting to break down, $k/M_*\rightarrow\mathcal{O}(1)$. However, the region of parameter space for which the phase transition completes increases if one includes the leading order backreaction from the stabilizing dynamics~\cite{Konstandin:2010cd}, and, to a lesser extent, if one reduces the IR scale below the TeV scale (as in our case)~\cite{Kaplan:2006yi}. Furthermore, in RS models with position-dependent curvature, as occurs for the (possibly more realistic) case of a warped throat~\cite{Klebanov:2000hb,Brummer:2005sh}, the phase transition completes with a markedly less-stringent constraint on $N$~\cite{Hassanain:2007js}.\footnote{Technically, the constraint is on the IR value of $N$; see the discussion in Appendix~\ref{app:phase_trans}.} We discuss this in more detail in Appendix~\ref{app:phase_trans}.

Note that, with respect to the above freeze-out calculations, it is not sufficient to simply demand that the phase transition to the RS-like phase completes at some temperature $T_n$. If this temperature is much smaller than the critical temperature, $T_n<T_c$, 
the universe enters a supercooled phase during which it undergoes inflation (since the energy density is dominated by the cosmological constant in the false vacuum). This would dilute the thermal relic density of DM. Our calculations provide a good approximation if the phase transition happens sufficiently fast with $T_n \simeq T_c$, as occurs for e.g.~the RS-like models studied in~\cite{Brummer:2005sh,Hassanain:2007js}, for which the geometry is $AdS_5$-like in the UV but increasingly differs from $AdS_5$ towards the IR. For our purposes it is sufficient to note that warped RS-like models exist for which the phase transition can successfully complete in the desired fashion. As we discuss in Appendix~\ref{app:DM_entopy}, for $T_n \simeq T_c$ the phase transition happens close to equilibrium and the total entropy density remains (approximately) conserved. This ensures that the DM density is not diluted and the above calculations are reliable.

Once the phase transition is complete and the universe is in the RS phase (equivalently, the confined phase), the energy density in the hidden sector consists of a gas of light KK modes (composites) with $\mathcal{O}(1)$ degrees of freedom. These KK modes cascade decay to lighter KK modes via two-body and higher-order decays on short timescales (set by the IR scale), provided $k/M_*$ is not hierarchically small (as discussed in Section~\ref{sec:hidden_kk_mode_pheno_one_throat}). Eventually one has a collection of $n=0$ (and possibly $n=1$) KK modes, which can only decay to the SM. These decays require a kinetic mixing insertion and therefore proceed over a longer timescale. For the zero mode, we have $\Gamma_0\sim \alpha c_W^2\epsilon_0^2 \Lambda_{\ir}/15$ and the condition $\Gamma_0\gtrsim H(T\approx \Lambda_{\ir})$ gives
\bea
\epsilon_0\ \gtrsim\ 10^{-8} \times\left(\frac{\Lambda_{\ir}}{\mathrm{GeV}}\right)^{1/2}.\label{zero_mode_decay_confine}
\eea
This condition, which can be more stringent than \eqref{sm_cft_equilibrium}, ensures that the lightest modes are in equilibrium with the SM after the phase transition. 

Dark radiation during big bang nucleosynthesis (BBN) must not contribute more than 10\% to the total energy density (see e.g.~\cite{Calabrese:2011hg}). Since the deconfined CFT typically has a large number of degrees of freedom (cf.~Eq.~\eqref{energydensity}), we require $\smash{\Lambda_\ir > \text{few MeV}}$ to ensure that the phase transition happens before BBN. Combined with Eq.~\eqref{zero_mode_decay_confine}, this ensures that the zero-modes decay before BBN. Alternatively, if Eq.~\eqref{zero_mode_decay_confine} is not satisfied, BBN leads to the following constraints on $\epsilon_0$: Decays of zero-modes to hadrons (i.e.~for $\smash{\Lambda_\ir \gtrsim \Lambda_{\text{QCD}}})$ around the time of BBN can lead to p-n interconversion and hadrodissociation and thereby change the primordial abundance of light elements. We thus require~\cite{Kawasaki:2004qu}
\be
\Gamma_0 \, \gtrsim \, 100 \text{ s}^{-1}  \quad \Rightarrow \quad \epsilon_0 \, \gtrsim \, 10^{-10} \, \sqrt{\frac{\text{GeV}}{\Lambda_\ir}} \, .
\label{BBNbound}
\ee
If only leptonic decay channels are kinematically allowed, the most stringent constraints arise from photodissociation of light elements and give $\smash{\Gamma_0 \, \gtrsim \, 1\text{ s}^{-1}}$. In this case, the bound on $\epsilon_0$ becomes correspondingly weaker. Note furthermore that, if Eq.~\eqref{zero_mode_decay_confine} is not satisfied, the zero-modes can potentially start dominating the energy density of the universe. Their eventual decay to the SM then leads to a new phase of reheating. As the zero-mode decays only to the SM and not to the DM, this would change the energy balance between the DM and the SM. A different annihilation cross section (compared to Eq.~\eqref{DMcs}) would then be required to obtain the right DM relic abundance. 
\section{Cosmic Rays from Dark Matter Annihilation \label{sec:cosmi_ray}}
Cosmic rays produced during DM annihilations and/or decays have received much attention in recent years in connection with experiments like INTEGRAL~\cite{Knodlseder:2003sv,Weidenspointner:2006nua}, PAMELA~\cite{Adriani:2008zr}, $Fermi$~\cite{Abdo:2009zk,Ackermann:2010ij} and others. It is clear that our Secluded Dark Matter model, in which DM annihilates into hidden-sector states that eventually decay to the SM, has the potential to produce observable cosmic-ray signals. The precise nature of the signal depends on the details of the model, like the values of $M_\dm$ and $\Lambda_{\ir}$, and on whether the DM is Dirac or Majorana. A full analysis of the cosmic ray signal for arbitrary $\Lambda_{\ir}$ is beyond the scope of this work. However, we will discuss some general properties of cosmic ray spectra in warped models of Secluded Dark Matter below, and explain why this framework is \emph{unlikely} to have any connection with the positron excess observed by PAMELA~\cite{Adriani:2008zr}, $Fermi$~\cite{Abdo:2009zk,Ackermann:2010ij} and others.\footnote{A warped model of DM with a single sub-TeV vector has been constructed in Ref.~\cite{Gherghetta:2010cq} in an effort to address the positron excess.} 
\subsection{General Features\label{subsec:general_cosmic}}
In conventional models of Secluded Dark Matter, the dominant annihilation channels are typically $t$-channel processes like $\psi\, \bar{\psi}\rightarrow2\gamma'$, or $s$-channel processes like $\psi\,\bar{\psi}\rightarrow \gamma' \, h'$ (the latter proceeding through an off-shell $\gamma'$). In the centre-of-mass frame of the DM (which coincides with the galactic rest frame, to good approximation), the hidden-sector final states have energies on the order of the DM mass and are thus highly boosted when $M_\dm\gg \Lambda_\hid$.\footnote{Here the hidden sector states like $\gamma'$ and the Higgs$'$ have $\mathcal{O}(\Lambda_\hid)$ masses.} These hidden sector states subsequently decay to the SM, producing cosmic-ray particles with kinetic energies of $\mathcal{O}(M_\dm)$. Adding a hidden CFT to the Secluded Dark Matter framework modifies the annihilation spectra in two important ways. Firstly, $s$-channel annihilations into the CFT produce CFT-states/KK-vectors with mass of $\mathcal{O}(M_\dm)$. Secondly, there are sub-dominant $t$-channel annihilations that produce light CFT-states/KK-vectors. We will outline how these new channels modify the cosmic ray spectrum.

Cosmic rays produced through $s$-channel annihilations can have a rather different injection spectrum compared to standard secluded DM models. As detailed in Section~\ref{sec:DAannihilations}, the dominant $s$-channel annihilation is $\psi\, \bar{\psi}\rightarrow X_n$, whereby the UV-localized DM creates a KK vector with mass $m_n\simeq 2M_\dm$ that escapes into the bulk. The production cross section is not sensitive to the details of the IR geometry, which is dual to the statement that the mass gap of the CFT is irrelevant at energies far above the confinement scale. The heavy KK vector decays to lighter KK modes, initiating a hidden sector cascade. Studies of the relevant KK wavefunction overlap-integrals reveal that these decays display an approximate KK number conservation~\cite{Csaki:2008dt,McDonald:2010fe}. This reflects an approximate conservation of momentum along the fifth dimension, due to the fact that the gravity description is weakly coupled. The consequence of the approximate KK number conservation is that the kinetic energy of the dominant decay products in each step of the cascade is typically $\lesssim\Lambda_\ir$~\cite{Hatta:2008tn,slowcomp,Strassler:2008bv,slowcomp1,slowcomp2,Csaki:2008dt,McDonald:2010fe}. The cascade continues until the lightest states in the warped sector are reached, leading to a large number of particles with mass and kinetic energy of order $\Lambda_\ir$. Using conservation of energy, the resulting number of light KK states is estimated to be of order $M_\dm/\Lambda_\ir$. These light states decay, in turn, to light SM fields with masses $m_{\sm}\lesssim\Lambda_\ir$.

Combining these elements, the $s$-channel production of cosmic rays can be schematically written as
\bea
\psi\bar{\psi}\longrightarrow X_{n\gg1} \longrightarrow  (M_{\dm}/\Lambda_{\ir})\times (X_{n\sim1}+X_0)\longrightarrow (M_{\dm}/\Lambda_{\ir})\times f\bar{f},
\eea
where $f$ is a light SM field. Thus the final state consists of a high multiplicity of light SM states when $M_\dm/\Lambda_\ir\gg1$. Furthermore, these states are relatively soft, with typical energies of order $\Lambda_\ir$, which is to be contrasted with the $\mathcal{O}(M_\dm)$ energies expected in conventional models. Note that if the radion is light, or if there is a light hidden Higgs, these states can also appear at the end of the hidden-sector decay cascades.

The second modification compared to conventional models of secluded DM results from the fact that the DM can now annihilate to a tower of KK-modes/composites in the $t$-channel:
\bea
\sigma_t(\psi\bar{\psi}\longrightarrow\mathrm{anything})\simeq\sum_{n,n'}\sigma_{nn'}= \sum_{n,n'} \sigma(\psi \bar{\psi}\longrightarrow X_n X_{n'})\ .
\eea
 Although this cross section is dominated by zero-mode production, $\sigma_t\ \simeq\ \sigma_{00}$, subdominant production of higher KK modes can still impact the cosmic ray signal. The heavier a final-state vector is, the less boosted it is in the galactic rest frame, leading to a \emph{softening} of the $t$-channel cosmic-ray spectrum relative to conventional secluded DM models. Noting that $\sigma_{nn'}\sim g_n^2 g_{n'}^2\times (\sigma_{00}/g_0^4)$, and that $g_n^2/g_0^2\sim 1/n$, one might naively expect that the cosmic rays resulting from the subdominant creation of higher KK modes will be negligible. However, a given KK mode with mass $m_n$ will, in general, produce $N_n^\sm\sim m_n/\Lambda_\ir\sim n$ light SM fields after decaying/showering in the hidden sector. Thus, for example, the leading term in the product $\sigma_{0n}\times N^\sm_n$ is approximately independent of $n$ (up to a subdominant logarithmic dependence, see Eq.~\eqref{DM_coupling_n}). 

Denoting the number of SM states produced in the annihilation $\psi \bar{\psi}\rightarrow X_n X_{n'}$ by $N_{nn'}^\sm$, one can estimate the number of SM fields produced in $t$-channel annihilations to higher KK modes, relative to those produced by the dominant $t$-channel process ($\psi \bar{\psi}\rightarrow 2 X_0$), as
\bea
N^\sm_t\ =\ \frac{\sum_{n,n'}\sigma_{nn'}N_{nn'}^\sm + 2\sum_n \sigma_{0n}N^\sm_{0n}}{\sigma_{00}N^\sm_{00}},\label{lepton_ratio}
\eea
where the sums run over $n,\, n'>0$. Using $\sigma_{nn'}\sim g_n^2 g_{n'}^2\times (\sigma_{00}/g_0^4)$,  the coupling relations~\eqref{DM_coupling_n}, and converting the sums to integrals, one can readily estimate \eqref{lepton_ratio}. For example, with $k/M_*=0.1$, $M_\dm\sim$~TeV, and $\Lambda_\ir\sim$~GeV, one obtains $N^\sm_t\sim15$ SM fields for every SM field created via boosted zero-mode production. Clearly, despite the suppression of the higher-mode coupling constants, the production of higher modes can produce a significant softening of the cosmic ray spectrum.

In summary, we expect the $s$-channel to produce $\mathcal{O}(M_\dm/\Lambda_\ir) \gg 1$ light composites per annihilating DM pair, with kinetic energies of order $\Lambda_\ir$. These states eventually decay to the SM, producing cosmic rays with kinetic energies on the order of the confinement scale. If present, the $s$-channel typically dominates the $t$-channel by a factor $\sim N$, so that these softer cosmic rays are the dominant effect. This leads to a ``bump'' in the cosmic ray spectrum at low energies of order $\Lambda_\ir$ that is not present in conventional models of secluded DM. Furthermore, the $t$-channel creates both boosted SM fields with energies of $\mathcal{O}(M_\dm)$, and a number of softer cosmic-ray particles due to the subdominant production of higher KK-modes. Alternatively, in the case that the fermion has a small mass-split, the present day abundance is comprised only of the lightest state $\psi_1$. Then the $s$-channel process is not present (or, more specifically, highly suppressed), while the $t$-channel annihilations remain. The cosmic rays spectrum is modified accordingly due to the absence of the large number of soft states with $\mathcal{O}(\Lambda_\ir)$ energies. 

Note that unstable SM particles will subsequently decay, and the final state ultimately consists of electrons/positrons, neutrinos, photons and/or (stable) nuclei. The stable particles produced by DM annihilations will propagate through the galaxy and may be detected on earth. Electrically neutral particles -- photons and neutrinos -- travel on straight lines and their flux at the earth is straightforward to determine by line-of-sight integrals. The charged electrons/positrons and nuclei, on the other hand, are deflected by the magnetic field in the galaxy, and their flux at the earth is determined by the diffusion-loss equation which governs their dynamics.

Though it is beyond the scope of this work to perform a detailed analysis of the cosmic ray signal, in the following section we will generate an example cosmic ray injection spectrum in order to demonstrate the effects of higher KK mode production and $s$-channel annihilations. It would be interesting to study the cosmic ray signal in more detail to see if parameter space exists for which the presence of the CFT permits an observable modification of the cosmic ray spectrum.
\subsection{Injection Spectrum of Cosmic Rays from DM Annihilation\label{sec:InjectionSpectrum}}
In order to calculate the flux of cosmic rays at the earth, one must determine the kinetic-energy spectrum for stable SM particles produced by the annihilation of a DM pair, i.e.~the number density of particles per unit energy $\smash{d N/d E}$. 
As should be clear from the preceding sections, the mechanisms by which the spectrum is
generated, which include hidden sector cascades, are somewhat
complicated. Nonetheless we can obtain some quantitative
approximations for the gross structure of the expected spectrum. This spectrum consists of two distinctive contributions, due respectively to the $s$-channel and $t$-channel annihilation of DM, which we discuss separately in the next two subsections. 
\subsubsection{Contribution from the $\mathbf{s}$-channel}

Let us assume that we have determined (e.g.~using Monte Carlo programs) the spectrum of a SM particle species $i$ resulting from the decay of a light composite,
\be
\frac{dN^{(i)}_0}{dE_0}\, ,
\label{eq: zero-mode spectrum}
\ee
where $N^{(i)}_0$ is the average number of particles $i$ resulting from the decay of \emph{one} composite and $E_0$ their kinetic energy in the rest frame of the parent particle. Given this spectrum, we now estimate the spectrum of particles $i$ in the galactic rest frame that results from the annihilation of one DM pair. To this end, we have to boost the spectrum to the galactic rest frame and take the multiplicity of decay products into account. 

The composites, whose decay produces the SM particles $i$ with spectrum $dN^{(i)}_0/dE_0$, result from the $s$-channel annihilation of the DM and the ensuing cascade in the CFT. As discussed in the last section, their average kinetic energy is thus of order $\Lambda_\ir$ in the galactic rest frame. As the theory is strongly coupled, however, it is difficult to obtain a quantitative statement about the distribution of these kinetic energies from the gauge-theory point of view. Similarly, unless  $k/M_*$ is extremely small, the higher KK modes are
strongly coupled too (see Eq.~\eqref{width_parama} for large $n$), limiting the utility of the gravity point of view. In absence of a probability distribution for the momenta of the light composites, we will content ourselves with discussing a limiting case: We shall assume that the radion is somewhat lighter than the IR scale or that the warped sector contains a light Higgs. In either case, as discussed in Sect.~\ref{sec:hidden_kk_mode_pheno_one_throat}, all heavier KK modes decay to these states and vector-zero modes before the decays to the SM. If the distribution of the momenta of these light states is relatively narrow around the average value $\langle p \rangle \sim \Lambda_\ir$, the boost to the galactic rest frame takes the form of a simple convolution.\footnote{
Consider scalar fields $\phi_i$ with masses $m_i$ which undergo cascade decays $\phi_2 \rightarrow \phi_1\phi_1$ and subsequently $\phi_1 \rightarrow \phi_0 + X$. Let us denote the spectrum of $\phi_0$  by $dN/dx_0$, where $x_0\equiv 2 E_0 /m_1$ and $E_0$ is the energy in the rest frame of $\phi_1$. We want to determine the spectrum in the rest frame of $\phi_2$. As discussed in Appendix A in~\cite{Mardon:2009rc}, in the limit $m_2 \gg m_1$, the boost into this frame takes the form of a simple convolution:
$$ \frac{d N}{d x_1} = \int^1_{x_1} \frac{dx_0}{x_0} \frac{dN}{dx_0} + \mathcal{O}\left(\frac{m_1^2}{m_2^2}\right) \, ,$$
where $x_1\equiv 2 E_1 /m_2$ and $E_1$ is the energy in the rest frame of $\phi_2$. If the light composites at the end of the cascade have momenta in the galactic rest frame with a relatively narrow distribution around the average value $\langle p \rangle$, then we can think of them as originating from the two-body decay of a particle with mass $2 \langle p \rangle$. Accordingly, if $ \langle p \rangle$ is much larger than the mass of the light composites, we can apply the above convolution to boost into the rest frame of this `parent particle' and thus the galactic rest frame.
\label{footnote: convolution}
} Taking the multiplicity of final states into account, the spectrum of particles $i$ can then be estimated as
\be
\frac{d N^{(i)}_s}{d E} \approx  \Theta(1-x_\ir) \frac{M_\dm}{\Lambda_\ir^2} \int^1_{x_\ir} \, \frac{d x_0}{x_0} \, \frac{dN^{(i)}_0}{dx_0} \, ,
\label{eq: s-channel spectrum}
\ee
where $x_\ir \equiv E / \langle p \rangle$ and $E$ is the kinetic energy in the galactic rest frame. Furthermore, $x_0 \equiv E_0 / m_0$, where $m_0 \ll \Lambda_\ir$ is the mass of the particle -- either a vector zero-mode, a radion or a Higgs -- whose decay produces the spectrum $dN^{(i)}_0 / dE_0$. The $\Theta$-function takes into account that, if the distribution of momenta is relatively narrow, the maximal kinetic energy of SM particles resulting from the $s$-channel is $\langle p \rangle$. If the distribution is broader, the spectrum is somewhat softer than Eq.~\eqref{eq: s-channel spectrum}.
\subsubsection{Contribution from the $\mathbf{t}$-channel}

The $t$-channel allows the annihilation of the DM to pairs of KK vectors $X_n$ and $X_{n'}$. Let us first consider the process
\be
\psi+\bar{\psi}\rightarrow X_0 + X_n \, ,
\ee
where $n$ is arbitrary. For KK numbers $n$ such that the mass of the $n$-th KK vector in this process satisfies $m_n \ll M_\dm$, the zero-mode has a momentum $|\vec{p}|\simeq M_\dm$ in the galactic rest frame. 
When it decays, it accordingly produces particles with momenta of this order. As before (cf.~Eq.~\eqref{eq: zero-mode spectrum}), we denote the spectrum of SM particles $i$ arising from this decay by $dN^{(i)}_0 / dE_0$, where $E_0$ is the energy in the rest frame of the parent composite. Since ${m_0 \ll M_\dm}$, the boost into the galactic rest frame is a simple convolution (see footnote \ref{footnote: convolution}) and we find
\be
\frac{dN^{(i)}_{t,0}}{dE } \, = \,  \Theta(1-x_\dm) \frac{1}{M_\dm}  \int^1_{x_\dm} \, \frac{d x_0}{x_0} \, \frac{dN^{(i)}_0}{dx_0} \, ,
\label{eq: t-channel spectrum1}
\ee
where $x_\dm \equiv E / M_\dm$. 

Let us next consider the process 
\be
\psi+\bar{\psi} \rightarrow X_{n} + X_{n'} \, ,
\label{eq: DM+DM->X_l+X_n}
\ee
with $n>0$ and $n'$ again arbitrary. The $n$-th KK mode in this process decays to lighter particles, producing $\mathcal{O}(m_n/\Lambda_\ir)$ light states at the end of the cascade (this number follows as before from energy conservation since said light states have energies of order $\Lambda_\ir$) which in turn decay to the SM. In order to obtain the resulting spectrum, we can make use of the results from the last section: One can think of the $s$-channel annihilation of the DM as producing a vector KK mode with mass $m_n \simeq 2 M_\dm$ which is at rest in the galactic rest frame (and which subsequently cascades to lighter KK modes). With the replacement $2 M_\dm \rightarrow m_n$ in the spectrum in Eq.~\eqref{eq: s-channel spectrum}, we therefore obtain the spectrum from the cascade of the $n$-th KK mode, given in \emph{its} rest frame.
Let us restrict to processes with KK numbers such that $m_n,m_{n'} \ll M_\dm$. The KK modes, which are initially produced from the $t$-channel annihilation, then carry momenta $|\vec{p}|\simeq M_\dm$ in the galactic rest frame. Since the momenta of their decay products relative to each other are small (of order $\Lambda_\ir$), the momentum of the parent KK mode is equally distributed during the cascade. The $\mathcal{O}(m_n/\Lambda_\ir)$ light states that emerge at the end of the cascade accordingly have momenta of order $\Lambda_\ir M_\dm / m_n$ in the galactic rest frame. The boost of the spectrum to the galactic rest frame takes again the form of a simple convolution (see footnote \ref{footnote: convolution}) and we find
\be
\frac{d N^{(i)}_{t,n}}{d E} \approx  \Theta(1-x_n) \frac{m_n^2}{M_\dm \Lambda_\ir^2} \int_{x_n}^1 \frac{d x_\ir}{x_\ir} \int^1_{x_\ir} \, \frac{d x_0}{x_0} \, \frac{dN^{(i)}_0}{dx_0} \, ,
\label{eq: t-channel spectrum2}
\ee
where $x_n\approx (m_n E) / (\Lambda_\ir M_\dm) \approx n\times( E  / M_\dm)$. 

In order to obtain the total spectrum due to the $t$-channel annihilation of DM, we sum over all KK modes:
\be
\frac{d N^{(i)}_{t}}{d E} = \sum_{n,n'=0} BR_{n,n'} \,\left( \frac{d N^{(i)}_{t,n}}{d E} + \frac{d N^{(i)}_{t,n'}}{d E} \right) = 2 \, \sum_{n=0} BR_n \, \frac{d N^{(i)}_{t,n}}{d E}\, ,
\label{eq: t-channel spectrum}
\ee
where $BR_{n,n'}$ is the branching fraction for the annihilation to $X_n X_{n'}$ and ${BR_n \equiv \sum_{n'} BR_{n,n'}}$.
The upper bound in the sums, which we have not written, is set by the mass of the heaviest kinematically accessible KK mode in the process and is of order $M_\dm / \Lambda_\ir$. Note that the spectrum in Eqs.~\eqref{eq: t-channel spectrum1} and \eqref{eq: t-channel spectrum2} was derived under the assumption that $m_n,m_{n'} \ll M_\dm$ which is no longer fulfilled close to this upper bound. We will therefore cut the sum off at KK numbers $n' \approx 10^{-1}M_\dm / \Lambda_\ir$ and neglect the remaining terms (which only give a small correction). Using the results from Section~\ref{sec:DAannihilations}, we find for the branching fraction
\be
BR_n \approx 
\begin{cases}
1  &\text{for } n = 0\\
\frac{1}{20n} &\text{for } n \neq 0 \, .
\end{cases}
\ee

Let us finally combine the $s$-channel spectrum from the last section and the $t$-channel spectrum to obtain the total spectrum from the annihilation of the DM in our scenario:
\be
\frac{d N^{(i)}}{d E} = BR_s \, \frac{d N^{(i)}_{s}}{d E} +  BR_t \, \frac{d N^{(i)}_{t}}{d E} \, .
\label{eq:combinedspectrum}
\ee
This is the spectrum that enters into the line-of-sight integral or the diffusion-loss equation to determine the cosmic ray flux at the earth. Using Eq.~\eqref{relating_s_t_channel}, we find the branching fractions 
$BR_s$ and $BR_t$ for respectively the $s$-channel and the $t$-channel to be
\be
BR_s \approx 1 \qquad \qquad BR_t \approx \frac{g_5^2 k }{6 \pi^2} \, . 
\ee
\subsubsection{An Example}

For definiteness, let us assume that the radion and the Higgs (if present) have a mass in the range ${3 m_0 \gtrsim m_{h',r} > 2 m_0}$. They then decay into two on-shell, relatively slow (in their rest frame) vector zero-modes. Let us further choose $m_0 = 260$ MeV, allowing the vector zero-modes to decay into electrons and muons but not pions or heavier hadrons. 
The muons subsequently decay to electrons and neutrinos. Photons are produced as final state radiation and in radiative muon decays. For the resulting spectra of these stable SM particles, analytic formulae (approximate for photons) are available. Let us focus on the spectrum of electrons (or positrons). In the rest frame of the parent zero-mode, it reads (see \cite{Mardon:2009rc})
\be
\frac{dN^{(e)}_0}{dx_0} \, = \, BR_e\, \delta(1-x_0) \, + \,BR_\mu \left( \frac{5}{3} \, - \, 3 x_0^2 \, + \, \frac{4}{3} x_0^3 \right) \, ,
\ee
where $x_0 = 2 E_0 / m_0$ and $BR_e$ and $BR_\mu \simeq 1-BR_e$ are the branching fractions to electrons and muons, respectively. Using Eq.~\eqref{VectorsToSM}, we find $BR_e \simeq 1/2$ for $m_0 = 260$ MeV.

Let us first determine the contribution from the $t$-channel. The spectrum due to a zero-mode is obtained using Eq.~\eqref{eq: t-channel spectrum1} which gives:
\be
\frac{dN^{(e)}_{t,0}}{dE}  =  \Theta(1 - x_\dm)\, \frac{1}{M_\dm} \left[  BR_e\, + \, BR_\mu \left(-\frac{19}{18} \, + \, \frac{3}{2}x_\dm^2 - \frac{4}{9} x_\dm^3 \,-\, \frac{5}{3} \log x_\dm \right)\right] \, ,
\label{eq: ex:t0}
\ee
where $x_\dm = E / M_\dm$. Similarly, applying Eq.~\eqref{eq: t-channel spectrum2}, we find the spectrum due to higher KKs:
\begin{multline}
\frac{dN^{(e)}_{t,n}}{dE}  \approx \Theta(1-x_n)  \frac{n^2}{M_\dm}  \left[ - BR_e \, \log x_n  +  BR_\mu \left( \frac{65}{108}  -  \frac{3}{4} x_n^2   \right. \right. \\\left. \left.+  \frac{4}{27} x_n^3  + \frac{19}{18} \log x_n +  \frac{5}{6} \left(\log x_n\right)^2 \right)\right]  \, ,
\end{multline}
where $x_n \approx n E / M_\dm$. The total $t$-channel spectrum follows from Eq.~\eqref{eq: t-channel spectrum}. Separating out the zero-mode contribution and replacing the sum by an integral, we can approximate the resulting expression as
\begin{multline}
\frac{d N^{(i)}_{t}}{d E} \approx \frac{dN^{(e)}_{t,0}}{dE} + \frac{-60 \,  x_\dm^5+400 \, x_\dm^4 +700 -1000 \, x_\dm^2  + 900 \, x_\dm^2 \left(\log x_\dm- (\log x_\dm)^2 \right)}{4 \cdot 10^4 \, x_\dm^2 M_\dm} \, .
\label{eq: ex:th}
\end{multline}
The $s$-channel spectrum, finally, follows from Eq.~\eqref{eq: s-channel spectrum} (with $x_\ir \approx E / \Lambda_\ir$):
\be
\frac{d N^{(i)}_s}{d E} \approx  \Theta(1-x_\ir) \frac{M_\dm}{\Lambda_\ir^2} \left[  BR_e\, + \, BR_\mu \left(-\frac{19}{18} \, + \, \frac{3}{2}x_\ir^2 - \frac{4}{9} x_\ir^3 \,-\, \frac{5}{3} \log x_\ir \right)\right] \, .
\label{eq: ex:s}
\ee

\begin{figure}[t]
\centering
\includegraphics[width=11cm]{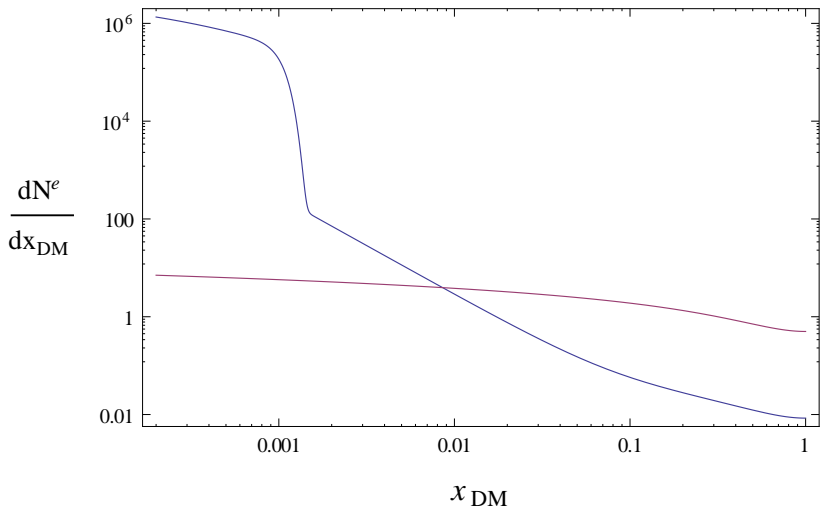} 
\caption{The electron injection spectrum $dN^{(e)} /dx_\dm$ plotted as a function of $x_\dm=E/M_\dm$ for our model of Secluded Dark Matter (blue curve). The `bump' at low energies (i.e.~small $x_\dm$) is due to the $s$-channel. For comparison, we include the injection spectrum for a conventional scenario of Secluded Dark Matter (red curve) where the only available annihilation channel is $\bar{\psi}\psi\rightarrow2\gamma'$.}\label{Fig:InjSpec}
\end{figure}

We plot the total spectrum, consisting of the contributions in Eqs.~\eqref{eq: ex:th} and \eqref{eq: ex:s} combined according to Eq.~\eqref{eq:combinedspectrum}, for the case $M_\dm=1$ TeV, $\Lambda_\ir \approx 1$ GeV (corresponding to ${m_0 = 260 \text{ MeV}}$) and $g_5^2k=1$ in Fig.~\ref{Fig:InjSpec}.\footnote{These values are consistent with the condition \eqref{DM_CFT_therm_condition} for $k/M_*\rightarrow1$.} For comparison we also plot the spectrum 
for a conventional Secluded DM scenario, in which case the $t$-channel process to two vectors with mass $260 \text{ MeV}$ is the only available annihilation channel. Integrating the spectrum over all energies, we find that there are $\sim 15$ leptons created
by $t$-channel annihilations to higher KK modes for each boosted lepton created from zero-mode production. The low-energy spectrum is dominated by the $s$-channel, which contributes $\sim 1000$ leptons for each annihilating DM pair. As can be seen from Fig.~\ref{Fig:InjSpec}, the additional contributions from higher KK-mode production lead to a softening of the spectrum and to a marked increase at low energies, relative to the standard scenario. 

Let us emphasize that, if the $s$-channel is not available in the present-day due to a small DM mass-split, the low-energy bump in the spectrum will not appear. Also note that although the spectrum in Fig.~\ref{Fig:InjSpec} was generated under the assumption of a light radion and/or hidden-Higgs, if these scalars are not light/present the spectrum will retain the same qualitative features. As mentioned already, the assumption of a light-scalar was employed purely to simplify the analysis as it enables  a scenario where all the lightest hidden-sector states decay purely to electrons and muons.
\subsection{A Secluded CFT and the Cosmic-Lepton Excess \label{subsec:pamela}}
Having detailed the general features of the cosmic ray spectrum produced in our model, let us now explain why the model is unlikely to be connected to the cosmic-lepton excess observed by PAMELA/$Fermi$ and others. The standard picture invoked to explain this positron excess, seen at energies $\gtrsim10$~GeV, involves TeV-scale DM annihilating into $\mathcal{O}(\mathrm{GeV})$ mediators~\cite{ArkaniHamed:2008qn}. The resulting boosted mediators then decay to produce boosted light SM fields (like electrons). The present-day annihilation cross section necessary to produce the signal is related to the cross section in the early universe by a boost factor:
\bea
\langle \sigma v\rangle_{\pd}\ \simeq\ BF\times  \langle \sigma v\rangle_{\eu},\label{standard_bf}
\eea
where $BF\gtrsim 10^2$ is the conventional boost factor (attributed to the Sommerfeld enhancement), and $\langle\sigma v\rangle_{\eu}\approx 3\times 10^{-26}~\mathrm{cm}^3/\mathrm{s}$ is the standard cross section for a WIMP. One of the greatest hurdles for these models is to generate a sufficiently large boost factor while remaining consistent with experimental constraints from, e.g.,  the cosmic microwave background, and gamma-rays from the galactic halo. Typically it can be difficult to remain consistent with the data and obtain values $BF\gtrsim 10^2$.

Naively it seems that for $M_\dm\sim$~TeV and $\Lambda_\ir\sim$~GeV our model has the properties needed to (potentially) generate the cosmic lepton excess --- particularly given the $\sim 20\%$ increase in the Sommerfeld enhancement due to CFT-composite exchange~\cite{McDonald:2012nc}. However, with the above information it is easy to see why this is unlikely to be the case. $t$-channel annihilations like $\psi\bar{\psi}\rightarrow 2X_0$ will produce boosted vectors that decay to light SM fields with energy of $\sim M_\dm$. Let us define the boost factor $BF_{\cft}$, which relates the cross section for boosted zero-mode/$\gamma'$ production in our secluded CFT model to that required in the present-day to generate the PAMELA/$Fermi$ signal:
\bea
\langle \sigma v\rangle_{\pd}\ \simeq BF_{\cft}\times  \langle \sigma_{00} v\rangle,
\eea
where $\sigma_{00}\simeq \pi\alpha_0^2/(M_\dm^2v)$ [see Eq.~\eqref{dm_ann_2X0}]. Noting that the cross section in the early universe is set by $\sigma_s\sim N\times\sigma_{00}$, and that $\langle \sigma_sv\rangle\approx \sqrt{g_\sm/g_*}\times\langle \sigma v\rangle_{\eu}$ [see Eq.~\eqref{DMcs}], one can relate $BF_\cft$ to the standard value:
\bea
BF_\cft\ &\sim&\ BF\times N\times \sqrt{\frac{g_\sm+\mathcal{O}(N^2)}{g_\sm}}\approx BF\times
\left\{
\begin{array}{cccl}
N^2&&\mathrm{for}&g_*\gg g_\sm\\
N&&\mathrm{for}&g_*\sim g_\sm\ .
\end{array}
\right.\label{boost_factor}
\eea
The requisite boost factor must therefore be larger than the standard WIMP value by a factor of $\mathcal{O}(N)$ to
$\mathcal{O}(N^2)$. This need for a larger boost factor is visible in Figure~\ref{Fig:InjSpec}, where the number of boosted leptons created via DM annihilations (the region with $x_\dm\sim\mathcal{O}(1)$) is seen to be suppressed relative to the standard WIMP result.

Given the well-known difficulty in generating values $BF\gtrsim 10^2$ while remaining consistent with constraints, the $\mathcal{O}(N-N^2)$ increase required in our model appears to exacerbate the tension with experimental constraints. However, one must account for the differences in the model when applying bounds on the boost factor. For example, one of the most robust constraints on the present-day DM annihilation cross section comes from WMAP5~\cite{Komatsu:2008hk}. Annihilations at the redshift of last scattering can affect the temperature and polarization angular power spectra of the CMB, leading to the constraint~\cite{Padmanabhan:2005es,Slatyer:2009yq}:
\bea
\langle \sigma v\rangle_\pd^{sat}\ <\ \frac{3.6\times 10^{-24}\mathrm{cm}^3/\mathrm{s}}{f}\left(\frac{M_\dm}{\mathrm{TeV}}\right)\, ,
\eea
where the superscript denotes that this bound applies to the saturated value of the Sommerfeld-enhanced low-velocity cross section, and $f<1$ parameterizes the fraction of DM annihilation energy that ionizes/heats the intergalactic medium~\cite{Slatyer:2009yq}.  Denoting the saturated Sommerfeld enhancement as $BF_{sat}$, the relevant cross section in the present case is approximately $BF_{sat}\times \langle \sigma_{00} v\rangle$. Using $\sigma_{00}\sim \sigma_s/N$ and $\langle\sigma_s v\rangle\approx \sqrt{g_\sm/g_*}\times\langle \sigma v\rangle_{\eu}$, gives
\bea
BF_{sat} \ \lesssim\ N\times\sqrt{\frac{g_*}{g_\sm}}\left(\frac{120}{f}\right) \left(\frac{M_\dm}{\mathrm{TeV}}\right)\, .\label{cmb_bound}
\eea
This shows that the CMB bound on $BF_{sat}$ is weaker by a factor of $ \mathcal{O}(N)$ when $g_*\sim g_\sm$, or by $\mathcal{O}( N^2)$ for $g_*\gg g_\sm$. Thus, one cannot assume that the standard bounds on the Sommerfeld enhancement directly apply in this case. Indeed, the weakening of the bound in \eqref{cmb_bound} with increasing $N$ matches the increase in the enhancement required by the lower value of the present-day (unenhanced) annihilation cross section; despite needing a larger present-day boost factor, the CMB bounds are no more severe.

Although the bounds need not be more severe, the greatest problem encountered when trying to achieve a larger value of the boost factor may result from the \emph{decrease} in the coupling constant required in the present model (for a given fixed value of $M_\dm$). Taking the DM mass fixed at $M_\dm\sim$~TeV to generate the high-energy lepton excess, the coupling constant required to achieve the correct DM abundance via freeze-out in the presence of a large-$N$ CFT is related to that of a standard WIMP as
\bea
\frac{\alpha_0}{\alpha_{\text{\tiny{WIMP}}}}\approx \frac{1}{\sqrt{N}}\ \left(\frac{g_\sm}{g_*}\right)^{1/4} \approx \left\{
\begin{array}{cccl}
1/N&&\mathrm{for}&\mathrm{large}~N\\
1/\sqrt{N}&&\mathrm{for}&~N\ \ \mathrm{such\ \ that}~~g_*\sim g_\sm \, . \ 
\end{array}
\right.  \label{coupling_small_for_cft}
\eea
The fine-structure constant must be smaller by a factor of
$\mathcal{O}(\sqrt{N})$ to $\mathcal{O}(N)$, depending on the value of
$N$. The size of the Sommerfeld enhancement depends
predominantly on two parameters, usually cast as
$\epsilon_\phi=m_0/(\alpha_0 M_\dm)$ and $\epsilon_v=v/\alpha_0$, such
that non-zero enhancements occur for $\epsilon_\phi,\, \epsilon_v
\lesssim1$~\cite{ArkaniHamed:2008qn}.\footnote{A non-zero mass splitting introduces a third parameter~\cite{Slatyer:2009vg}.} Furthermore, the enhancement increases for decreasing
values of $\epsilon_\phi$ and $\epsilon_v$, and off-resonant values of
the enhancement on the order of~$\gtrsim 10^3$ typically require
$\epsilon_\phi, \,
\epsilon_v\lesssim10^{-3}$~\cite{ArkaniHamed:2008qn}.\footnote{The
  boost factor required in the present framework is likely to be
  $\gtrsim 10^3$; for most of the benchmark points in Table~1 of
  Ref.~\cite{Finkbeiner:2010sm} an increase in the requisite local
  boost factor by $\sim 4$ is enough to necessitate values of
  $BF\gtrsim10^3$. }  These ingredients spell-out the primary
difficulty in trying to generate the PAMELA excess: According to
\eqref{boost_factor} the  requisite boost factor must be larger by a
factor of $\gtrsim N$, but the decrease in the fine-structure constant
by the factor $\lesssim 1/\sqrt{N}$ in \eqref{coupling_small_for_cft}
increases $\epsilon_\phi$ and $\epsilon_v$. For $m_0\sim 0.1-1$~GeV
and $M_\dm\sim$~TeV the required value of $\alpha_0$ typically forces $\epsilon_\phi>10^{-3}$ and thus forbids the successful realization of the larger boost factor.\footnote{We assume here that off-resonant values of the boost factor are employed, as it seems unlikely that one can successfully employ resonant values without a highly tuned set of circumstances.}

Note that in drawing these conclusions one must account for the increase in the Sommerfeld enhancement that results from exchange of the higher composites/KK-modes~\cite{McDonald:2012nc}. However, the $\gtrsim\mathcal{O}(1)$ decrease in the enhancement induced by the smaller values of $\alpha_0$ in \eqref{coupling_small_for_cft} typically overwhelms the $\sim20\%$ increase due to composite exchange. One possible exception to this statement is if the UV scale at which the theory becomes conformal is much less than the Planck scale, $k\ll M_{Pl}$. Then the increase in the enhancement due to composite exchange can approach $\mathcal{O}(1)$ values due to the increased coupling strength of the higher modes~\cite{McDonald:2012nc}. We do not consider this case here.

Together, these results make it unlikely that the PAMELA excess can be explained in models where the DM annihilates into a large-$N$ CFT. Said differently, the boosted cosmic leptons produced in the presence of a Secluded CFT are expected to, at best, comprise a subdominant component of the total cosmic lepton spectrum. Thus, in this framework, alternative sources of energetic leptons, presumably astrophysical in nature~\cite{Hooper:2008kg,Mertsch:2010fn}, are expected to explain the boosted-positron excess.

Note that we use the word ``unlikely'' rather than definitively rule
out models with Secluded CFT's as candidate explanations for the
lepton excess. The arguments above make the conservative assumption
that the local boost factor is not dominated by local-substructure
effects in the galactic halo. Should large amounts of local
substructure be present, the local DM velocity can be lower than the
average value for a smooth halo, thus generating a larger local boost factor. It has been argued that such substructure can allow models of annihilating DM to generate the lepton excess and remain compatible\footnote{Note that recent gamma-ray observations of the galactic center by HESS disfavor sizable regions of the parameter space, even with substructure~\cite{Abazajian:2011ak}, subject to some dependency on the nature of the DM density profile (the HESS galactic-center background subtraction method prevents the derivation of limits on DM annihilations when the inner $\sim450~pc$ of the Milky Way has a constant-density core~\cite{Abramowski:2011hc}).} with the most robust constraints~\cite{Slatyer:2011kg}. Indeed, in the split-fermion case with  $M_{\dm}\sim$~TeV and $\Lambda_\ir\sim$~GeV, our theory provides a concrete example of a model with ``new irrelevant annihilation channels''  according to the definitions of Ref.~\cite{Slatyer:2011kg}. However, we find that, even in the presence of large amounts of local substructure, the large-$N$ limit seems incompatible with the successful generation of the cosmic-lepton excess. This limit requires the coupling constant $\alpha_0$ to be smaller than that compatible with the results of Ref.~\cite{Slatyer:2011kg}. The small-$N$ limit does appear to be marginally consistent with their results, though only as $k/M_*\rightarrow\mathcal{O}(1)$, for which the perturbative gravity description begins to break down. Thus, we can safely say that, in the large-$N$ limit, the Secluded CFT model is not expected to generate a significant fraction of the observed boosted cosmic-lepton events. The ``smaller''-$N$ limit may be compatible with the results of Ref.~\cite{Slatyer:2011kg} if there are large amounts of local substructure. 
\subsection{Some Further Comments \label{subsec:comments}}
We have offered some general comments regarding the cosmic ray
spectrum and explained why the resulting spectrum is unlikely to be
connected to the PAMELA/$Fermi$ excess. More work is required to determine if
the resulting signal is compatible with the data for general values of
$M_\dm$ and $\Lambda_\ir$, particularly when $s$-channel annihilations
occur in the present-day. However, the above discussion is enough to
verify that the case of $M_\dm\sim$~TeV and $\Lambda_\ir\lesssim$~GeV
is compatible with existing cosmic ray constraints. This is certainly
true if the $s$-channel is absent in the present day, but likely also
holds if the $s$-channel is allowed. To see this, let us focus on the
case where the lightest hidden sector states decay directly to
electrons. Then, in the absence of the $s$-channel,  the cosmic ray
spectrum due to DM annihilations consists of a bump of
electrons/positrons at energies $\sim$~TeV, and a softer tail down to
energies of $\sim \Lambda_\ir$. The relevant (unboosted) production cross section
is less than the standard value, $3\times
10^{-26}\mathrm{cm}^3/\mathrm{s}$, by a factor of
$\mathcal{O}(N-N^2)$, so this spectrum is subdominant to the primary
source of the PAMELA/$Fermi$ excess (which is assumed astrophysical in
nature). We have undertaken some numerical investigations, employing a
standard approach to modeling cosmic ray propagation through the
galaxy to the earth, to verify that the softer states produced by higher
KK modes do not modify this statement.\footnote{In our investigations
  we accounted for the $\sim20\%$ increase in the Sommerfeld
  enhancement due to KK vector exchange but did not include the
  effects of substructure.} This case is therefore consistent with cosmic ray constraints.

If the $s$-channel is available in the present day, there is an
additional ``bump'' in the cosmic ray spectrum at energies of order
$\Lambda_\ir\lesssim$~GeV. However, this increase in the softer part of
the cosmic lepton spectrum is not of concern as it occurs in the
$\lesssim10$~GeV region of the spectrum, for which a large background
of positron sources exist and solar modulation significantly modifies
the spectrum, making the extraction of robust constraints more
difficult.

More generally we expect the Secluded DM model to be consistent with
cosmic ray constraints when the $s$-channel is absent in the present
day. In this case the present-day annihilation cross section is lower
than the standard WIMP value, and the cosmic ray signal is expected to
be below that expected in standard secluded DM models. Further study is
required to determine the viability of the signal when the $s$-channel
is available.
\section{Secluded Dark Matter via AdS/CFT\label{ads_cft}}
In this work we sought to study a model of Secluded Dark Matter in which the DM couples to a hidden CFT. After motivating the setup and sketching its features, we invoked the AdS/CFT correspondence to construct a  weakly-coupled dual 5D theory, with which our calculations were performed. This enabled us to avoid some of the perturbativity issues associated with the strongly-coupled conformal sector. In this section, after recapping some features of the 4D model, we discuss the correspondence in more detail.

The 4D model consists of a DM candidate ($\psi$) that is charged under a weakly-gauged global $U(1)'$ symmetry of a conformal hidden sector. In addition to permitting new annihilation channels and leading to potentially interesting phenomenology, the conformal sector confines in the IR in such a way that $U(1)'$  symmetry breaking is also triggered. This generates a radiatively-stable mass for $\gamma'$, with the symmetry breaking scale set by the mass gap of the CFT, $m_{\gamma'}\sim \Lambda_{\ir}$. Communication between the SM and the hidden sector proceeds through $\gamma'$, which mixes with hypercharge and thus acts as a ``mediator'' between the two sectors. The coupling to the (broken) CFT therefore provides a natural means to generate a stable mediator-scale via confinement. In addition to IR breaking via confinement, the conformal symmetry is (presumably) broken in the UV at some cutoff scale $\Lambda_{\uv}$. We focused on the case with $\Lambda_{\uv}\sim M_{Pl}$, though in principle the UV scale at which the theory enters the conformal regime could much be lighter.

Holographic arguments suggest that warped models on $AdS_5$ are dual to large-$N$ CFTs in 4D~\cite{Maldacena:1997re}. Interpreting the central charge of the CFT in terms of a large-$N$ $SU(N)$ gauge theory via $c=(N^2-1)/4$, the effective number of colors for the 4D dual of a pure-gravity theory is $N^2\simeq 16\pi^2(M_*/k)^3$~\cite{Maldacena:1997re,Gubser:1999vj}. RS models, possessing both UV and IR branes, are dual to
strongly-coupled 4D theories that are approximately conformal for energies $\smash{k\gg E \gg
R^{-1}}$~\cite{ArkaniHamed:2000ds}. The UV brane
is dual to an explicit breaking of conformal invariance in the 4D theory at high
energies due to a UV cutoff. The IR brane is dual to a further spontaneous breaking of conformal invariance at low energies. The spectrum of KK gravitons contains a massless zero-mode that is localized towards the UV brane. 
On the CFT side, this mode corresponds to a fundamental 4D graviton that is external to the CFT.\footnote{More generally, for a given bulk field $\mathcal{F}$ in the 5D theory, there is a fundamental field in the dual 4D theory that is determined by the UV-restriction $\mathcal{F}|_{UV}$. The UV-localized zero-mode graviton dominates the UV value of the bulk graviton field, $h^{5D}_{\mu\nu}|_{UV}\simeq h^{(0)}_{\mu\nu}|_{UV}$, so the fundamental massless graviton in the dual 4D theory corresponds ``mostly'' to the zero mode on the 5D side.} The 4D theory also contains a tower of spin-two composites, and the Goldstone boson of the spontaneously-broken conformal invariance (the dilaton), which, on the 5D side, are dual to the tower of KK gravitons and the radion, respectively. Note that for more general models on a slice of $AdS_5$ the value $N^2\simeq 16\pi^2(M_*/k)^3$ can be used as a guide for the number of colors in the dual theory. This is consistent with the interpretation that the 4D Planck mass is induced by CFT loops in the 4D theory, giving $M_{Pl}^2\sim \frac{N^2}{16\pi^2}\Lambda_{\uv}^2$, which should be compared with the 5D result of $M_{Pl}^2\simeq M_*^3/k$, after identifying $\Lambda_\uv\Leftrightarrow k$.

Our warped model also contains a bulk $U(1)_{\x}$ gauge symmetry. 
In the dual picture, the UV restriction of $X_\mu$ corresponds to a fundamental gauge boson (i.e.~external to the CFT) which weakly 
gauges a global $U(1)'$ symmetry of the CFT. This is analogous to the weakly-gauged global $U(1)_{em}$ symmetry of the QCD sector in the (low-energy) SM.\footnote{The analogy holds for the full weakly-gauged $\smash{SU(2)_L\times U(1)_Y}$ subgroup of the $\smash{SU(N_f)_L \times SU(N_f)_R \times U(1)_V}$ global symmetry in the QCD sector. It is suffice to consider the low-energy $SU(3)_c\times U(1)_{em}$ symmetry for our purposes.}  We denote the 4-vector potential for the fundamental gauge boson $\gamma'$ by $A'$ (with field strength $F'$).
The bare Lagrangian at the cutoff scale $\Lambda_\uv$ of the dual 4D theory then reads~\cite{Agashe:2002jx}
\bea
\mathcal{L}^{(\uv)}_{4D} \, \supset \, \mathcal{L}_{CFT}  +  \frac{1}{e'^{\, 2}_{\uv}} F'_{\mu \nu}  F'^{\mu \nu} \, + \, A'_\mu J_{CFT}^{\mu} \, ,
\eea
where $J_{CFT}^{\mu}$ is the conserved current of the global $U(1)'$ that is weakly gauged. 

The CFT contributes to the running of the gauge coupling from its value $e'_{\uv}$ at the UV scale down to lower energies. Assuming that the CFT has $N\gg1$ colors and $N_f\sim \mathcal{O}(1)$ flavors, the number of flavors for the $U(1)'$ sector is $N_f\times N \sim N$ and CFT color acts as $U(1)'$ flavor.  Running the coupling from the UV scale $\Lambda_\uv$ down to a momentum scale $p$ yields
\bea
\frac{1}{e'^{\, 2}(p)}  \sim  \frac{1}{e'^{\, 2}_{\uv}}+ \beta_{CFT}\left[ \ln(\Lambda_\uv/p) +\mathcal{O}(1)\right] \, ,\label{rge_4d}
\eea
where $\beta_{CFT}\sim N$. Note that $\beta_{CFT}$ is not completely determined by the gauge group representations as it depends on the strong dynamics of the CFT sector~\cite{Agashe:2002jx}. However, the conformal symmetry fixes the $p$-dependence in Eq.~\eqref{rge_4d}. Below the scale $\Lambda_\ir$, the 4D theory confines and the CFT sector ceases to affect the running, thus fixing the IR gauge-coupling at $e'_{\ir}=e'(\Lambda_{\ir})$. Compare this with the gauge coupling of the zero mode obtained with the dual RS model,
\bea
\frac{1}{g_{0}^2} \simeq \tau_\uv+\frac{1}{ g_5^2k}\left[\ln(kR)-\gamma\right] \, ,
\eea
where $\tau_\uv$ is the prefactor of a UV-localized kinetic term for the bulk gauge boson. We can identify the IR coupling on the CFT side with the zero-mode coupling, $\smash{e'_{\ir}\Longleftrightarrow g_0}$, if the number of flavors for $U(1)'$ on the CFT side is dual to the quantity $(g_5^2k)^{-1}$, i.e.~$N\sim (g_5^2k)^{-1}$. More generally, for $\tau_{\uv}=0$ we can identify the 4D running coupling as
\bea
\frac{1}{e'^{\, 2}(p)}  \Longleftrightarrow \frac{1}{ g_5^2k}\left[ \ln(2k/p)-\gamma\right] \, .\label{coupling_duality}
\eea
Note that the UV coupling $e'_{\uv}$ is determined by $\tau_{\uv}$ and, absent a UV-localized kinetic term (as assumed in the text), the gauge boson in the dual $4D$ theory is not dynamical at the scale $\Lambda_\uv$. Nevertheless, a kinetic term is induced by the running due to CFT loops, which is dual to the UV-to-UV value of the bulk-vector propagator on the RS side~\cite{PerezVictoria:2001pa,Agashe:2002jx}.

After confinement the 4D theory contains a tower of composite spin-one states ($\rho_n$). These states are (approximately) dual to the light KK modes of the bulk gauge boson in our RS model. 
We break the corresponding bulk $U(1)_{\x}$ symmetry either by a Higgs localized towards the IR or by imposing a Dirichlet BC on the IR brane. This is
dual to the spontaneous breaking of $U(1)'$ either by an explicit composite Higgs or in a way similar to Technicolor models. 

The warped model also contains a DM fermion $\psi$ and the SM sector, both localized on the UV brane. These correspond to fundamental fields in the dual 4D description (external to the CFT). The relevant part of the 4D Lagrangian at the cutoff scale $\Lambda_\uv$ reads
\be
\mathcal{L}^{(\uv)}_{4D} \, \supset \, \mathcal{L}_{SM} + \bar{\psi} \partial_\mu \gamma^\mu \psi + i \bar{\psi} A'_\mu \gamma^\mu \psi + \epsilon' F'_{\mu \nu} B^{\mu \nu}\, ,
\ee
where $B^{\mu \nu}$ denotes SM hypercharge. Note that this Lagrangian contains no direct coupling between the DM and the CFT, though the two sectors communicate via $\gamma'$ exchange.  Similarly, the mixing between $\gamma'$ and hypercharge induces a coupling between hypercharge and the CFT.

The fundamental field $\gamma'$ is determined by the UV value of the bulk gauge field $X_\mu$, which, as we have seen, is dominated by the zero mode. Therefore $\gamma'$ corresponds ``mostly'' to the zero-mode gauge boson~\cite{Batell:2007jv}. This allows us to understand the dominant $t/u$-channel DM annihilations in the dual picture approximately as annihilations into the fundamental gauge boson $\gamma'$,
\bea
\bar{\psi}\ \psi \longrightarrow X_0\ X_0 \qquad\Longleftrightarrow \qquad \bar{\psi}\ \psi \longrightarrow   \gamma'\ \gamma'.
\eea
Note that this production occurs locally on the UV brane in the 5D theory, and therefore
occurs
within the fundamental sector of the dual 4D theory. 

On the other hand, DM annihilations in the $s$-channel produce hidden
CFT states when $M_{\dm}/\Lambda_{\ir}\gg1$. These require a kinetic mixing insertion between $\gamma'$
and $\rho_n$ and proceed through an off-shell $\gamma'$. In the 5D
picture this corresponds to the local production, on the
UV brane, of a bulk vector that propagates into the bulk:
\bea
(\bar{\psi}\ \psi \longrightarrow\ Hidden)_{s-channel}   \qquad\Longleftrightarrow \qquad \bar{\psi}\ \psi \longrightarrow  CFT\ states.
\eea
As the DM mass is much greater than the confinement scale $\Lambda_{\ir}\sim
R^{-1}$, the hidden-sector modes produced in the $s$-channel are dual
to the continuum of RS2-like modes. Once produced, these states shower
within the hidden sector until the invariant mass of the decay
products is on the order of
the CFT breaking scale $\Lambda_{\ir}$, at which point the decay products are
a collection of narrow small-$n$ composites $\rho_n$ and the fundamental vector $\gamma'$. The lightest such modes
cannot decay within the hidden sector and therefore decay to light
SM fields via the induced mixing with hypercharge. The overall annihilation process may be schematically written as
\bea
\psi\bar{\psi}\longrightarrow CFT \longrightarrow  (M_{\dm}/\Lambda_{\ir})\times (\rho_{n\sim1}+\gamma')\longrightarrow (M_{\dm}/\Lambda_{\ir})\times f\bar{f},
\eea
where $f$ denotes a light SM field. As the kinetic energy of the decay products in each step along the hidden-sector cascade is typically $\lesssim\Lambda_\ir$, the final state SM fields will have energies of $\mathcal{O}(\Lambda_\ir)$. Thus, energy conservation mandates that there will be $\sim M_\dm/\Lambda_\ir$ final-state SM fields. 

Note that the cascading feature is readily understood from the dual gauge-theory point of view. To see this, consider the QCD process $q\bar{q}\rightarrow gluon \rightarrow q\bar{q}$. At energies far above the QCD scale $\Lambda_\text{QCD}$, the QCD coupling is small and the emission of energetic partons by the outgoing quark pair is suppressed. Instead, the quarks predominantly emit soft and collinear partons that evolve into collimated jets of particles. In the CFT dual of an RS model, on the other hand, the 't-Hooft coupling $\lambda$ is large on all energy scales. The vertex for radiating off partons in the $s$-channel annihilation process, $DM+ DM \rightarrow CFT\  partons$, is thus never suppressed. Accordingly, one expects the outgoing partons to successively branch into larger numbers of partons, distributing the energy approximately equally among all particles~\cite{slowcomp}. This evolution continues until the average energy of the partons reaches the confinement scale, leading to a large number of relatively slow composites, with mass and kinetic energy on the order of the confinement scale. 

We can also use the duality to derive the DM annihilation cross sections obtained via the 5D picture in Section~\ref{sec:DAannihilations}. For example, the $s$-channel production of CFT modes in the 4D picture proceeds through an off-shell $\gamma'$ and has the cross section
\bea
\sigma(\psi\bar{\psi}\rightarrow CFT)\sim N\times \frac{e'^4(2M_{\dm})}{M_{\dm}^2},
\eea
where the running coupling is evaluated at $p=2M_{\dm}$. Using the relation \eqref{coupling_duality} and the fact that $N\sim 1/g_5^2k$, we obtain the cross section for the dual 5D theory
\bea
\sigma(\psi\bar{\psi}\rightarrow CFT) \Longleftrightarrow \frac{g_5^2k}{\left[ \ln(k/M_{\dm})-\gamma\right]^2} \times \frac{1}{M_{\dm}^2}\sim\sigma_s(\psi\bar{\psi}\rightarrow X_n),
\eea
in agreement with the result found in Section~\ref{sec:DAannihilations}. Similarly, the $t$-channel calculation gives
\bea
\sigma(\psi\bar{\psi}\rightarrow \gamma'\gamma')\Longleftrightarrow  \frac{(g_5^2k)^2}{\left[ \ln(kR)-\gamma\right]^2} \times \frac{1}{M_{\dm}^2}\sim \sigma_t(\psi\bar{\psi}\rightarrow X_0X_0).
\eea
Observe that the 4D picture explains the $\mathcal{O}(g_5^2k)$ suppression of $\sigma_t$ relative to $\sigma_s$ found via the 5D picture. This reflects the fact that, in the 4D theory, the dual of  $\sigma_s$ depends on the number of colors in the CFT, while the dual of $\sigma_{t}$ involves fundamental fields and is therefore independent of $N$, giving
\bea
\frac{\sigma_t(\psi\bar{\psi}\rightarrow X_0X_0)}{\sigma_s(\psi\bar{\psi}\rightarrow X_n)}\ \sim\ \mathcal{O}(g_5^2k)\ \Longleftrightarrow\ \sim\frac{1}{N}.
\eea

\section{Some Comments and Future Directions\label{sec:comm}}
Before concluding we briefly comment on a few interesting modifications and
potential applications of our results.
\subsection{A Secluded Small-$N$ CFT}
We have studied a model of Secluded Dark Matter with the hidden-sector
gauge-symmetry $SU(N)' \times U(1)'$. Our analysis made use of the
AdS/CFT correspondence and a calculable dual RS-model was
considered. Such models are dual to 4D CFTs where the effective number
of colors is approximated by $N^2\simeq 16\pi^2(M_*/k)^3$. Validity of the RS
gravity description requires $M_*\gtrsim k$, so the dual theory is a
large-$N$ CFT.  Despite offering the benefit of admitting a calculable
dual 5D-theory, the large-$N$ limit does not encompass the most
general set of possibilities. An obvious alternative is that the
mediator scale is generated by a confining small-$N$ CFT.\footnote{These have been studied in the context of electroweak symmetry breaking~\cite{Luty:2004ye}.} The absence of a weakly-coupled dual description means a hidden-sector with a strongly-coupled small-$N$ CFT cannot be easily studied. However, we can make some general statements regarding the differences between  the small-$N$ and large-$N$ cases. 

Taking the Secluded Dark Matter to be charged under a weakly-gauged
global $U(1)'$ symmetry of a small-$N$ CFT, one retains $t$-channel
annihilations to $\gamma'$ with cross section $\smash{\sigma_t\sim
  e'^4/M_{\dm}^2}$. The cross section for $s$-channel annihilations
into the CFT  still goes like $\smash{\sigma_s\sim (N_f N)\times
  e'^4/M_{\dm}^2}$. Both channels eventually produce SM fields, though
the $s$-channel first undergoes a hidden-sector shower. The small-$N$
limit does, however, result in two key differences. With $N\sim
\mathcal{O}(1)$, the number of flavors in the $U(1)'$ sector is
approximately $N_f\times N\sim N_f$, and for $N_f\sim \mathcal{O}(1)$
the $s$-channel and the $t$-channel can now be related by an
$\mathcal{O}(1)$ factor. Also, the number of degrees-of-freedom in the
early hot-plasma need not be dominated by the CFT sector. In the large-$N$ limit we had $g_{\hid}\gg g_{\sm}$, giving $\smash{g_*= g_{\sm}+g_{\hid}\sim g_{\hid}}$, whereas the small-$N$ limit allows $g_{\hid}\lesssim g_{\sm}$ so that $g_*\sim g_{\sm}$. Combined, these results imply that the small-$N$ limit would allow thermal annihilation cross sections in the early universe that are on the order of the standard WIMP value, $\smash{\sigma\sim e'^4/M_{\dm}^2}$, and therefore the coupling $e'$ can be on the order of the usual size for WIMP models, $e'\sim g_{\weak}$. Indeed, one expects that theories with small-$N$ CFTs should map the theory space between the large-$N$ limit we have studied and the usual WIMP/Secluded-Dark-Matter models, resulting in lighter DM (for fixed coupling) or smaller couplings (for fixed DM mass).

These features may make the small-$N$ regime of more interest for the
PAMELA high-energy lepton signal. For $\sigma_s\sim \sigma_t$ the
boost factor is no longer required to be larger than the conventional
values (see e.g.~\cite{Finkbeiner:2010sm}). In addition, relative to conventional models,
extra boosted leptons will be created in the
channels $\sum_n\sigma_{0n}$, and the showering of
higher CFT states softens the spectrum, which is known to help alleviate
bounds from $\gamma$-ray production~\cite{Meade:2009rb}. In combination with the increased boost factor due to CFT composite exchange~\cite{McDonald:2012nc}, these ingredients would appear to make the small-$N$ limit of greater interest with regards to the high-energy lepton signal.

\subsection{Hidden Sector Showering and INTEGRAL}
The region of parameter space with $\Lambda_{\ir}\sim10$~MeV is potentially interesting. We have shown that $s$-channel annihilations produce a collection of $\sim M_{\dm}/\Lambda_{\ir}$ soft SM fields with mass $m_{\sm}<\Lambda_{\ir}$. This feature could be useful in relation to the $511$~keV photon excess observed by INTEGRAL~\cite{Knodlseder:2003sv,Weidenspointner:2006nua}. A candidate explanation for these photons involves the decay of an excited DM state to produce positrons with order MeV energies~\cite{Finkbeiner:2007kk}. The positrons then annihilate to produce the observed 511~keV photons. A mass gap of order $\Lambda_{\ir}\sim10$~MeV would provide a new way to generate soft positrons as a result of DM annihilations. DM candidates of mass $M_{\dm}$ would annihilate to produce CFT modes that shower within the hidden-sector. Upon reaching the mass gap, the products of the showering form hidden-sector composites with masses $\sim\Lambda_{\ir}$, which eventually decay to light SM fields. This provides a simple way of generating many soft positrons from a single annihilation event --- typically one would expect $\sim M_{\dm}/\Lambda_{\ir}$ electrons per annihilation. Work is required to determine if this explanation is viable; the high-energy tail of the injection spectrum must be subdominant to the background spectrum in order to reproduce the sharp low-energy injection morphology of the 511~keV line.

\subsection{Spatially-Secluded (or Composite) Dark Matter}
We have concentrated on the parameter space with
$M_{\dm}/\Lambda_{\ir}\gg1$, but smaller values of
$M_{\dm}/\Lambda_{\ir}\sim \mathcal{O}(1)$ could also be of interest.  Although one could retain a UV-localized DM candidate (equivalently, a DM particle external to the CFT) for $M_{\dm}/\Lambda_{\ir}\sim \mathcal{O}(1)$, it might be more sensible to consider the DM as part of the composite sector in this case.  Then the DM mass is  naturally related to the scale $\Lambda_{\ir}$, connecting the scale of conformal symmetry breaking to both $M_{\dm}$ and the $U(1)'$ symmetry-breaking scale. Indeed, DM masses of $M_{\dm}\sim 1-10$~GeV are of particular interest in relation to the results from DAMA/LIBRA~\cite{Bernabei:2010mq} and CoGENT~\cite{Aalseth:2010vx}, and composite DM candidates can be employed in this context (see e.g.~\cite{Frandsen:2011kt}). The 5D dual of a large-$N$ CFT with composite DM would include an IR localized DM candidate in order to naturally generate $M_{\dm}\sim\Lambda_{\ir}$. This suggests a more literal interpretation of the ``Secluded'' Dark Matter --- the DM could literally be spatially secluded from the SM sector in the fifth dimension.

\section{Conclusion\label{sec:conc}}
Secluded Dark Matter models offer a variant on the standard WIMP
picture and can modify our expectations for hidden sector
phenomenology and detection. In this work we added a strongly-coupled
(broken) CFT to a minimal Secluded Dark Matter model comprised of a
$U(1)'$-charged DM candidate. This provided a technically natural
explanation for the hierarchically small mediator-scale, with
confinement in the hidden sector generating $m_{\gamma'}\sim
\Lambda_\ir$. In addition, the mediator coupled both the DM and the SM
to the CFT, allowing the SM to communicate with the extended
secluded-sector. We studied the way in which these ingredients modify
the thermal history of the early universe, due to ($i$) new DM
annihilation channels, ($ii$) a (potentially) large number of
hidden-sector degrees of freedom, and ($iii$) a hidden-sector phase
transition at temperatures $T\ll M_{\dm}$ after freeze-out. We found
that viable parameter space exists such that the secluded CFT is
compatible with constraints and the correct DM abundance is thermally
generated. Furthermore, relative to standard models of Secluded Dark
Matter, the model predicts distinct modifications to the low-energy
phenomenology and cosmic ray signals,  due to the increased number of
light (composite) states in the secluded sector. Directions for future study include a more detailed 
analysis of the cosmic ray signal when $s$-channel annihilations are available in the present-day, and the inclusion of the DM in the composite sector. 
\section*{Acknowledgements\label{sec:ackn}}
The authors thank B.~Echenard,
T.~Gherghetta, A.~Hebecker, T.~Jacques, A.~Mueller,
K.~Petraki, and T.~Schwetz for
useful discussions. BvH was supported by the Australian Research Council and thanks SLAC for hospitality 
while part of this work was completed.

\emph{Note Added}: As we were finishing this work
Ref.~\cite{Lees:2012ra} appeared in which the BABAR collaboration
search for light hidden scalars that decay via $h'\rightarrow
2\gamma'$. The signal, assumed due to $e^+e^-\rightarrow h'\gamma'$,
thus consists of six-lepton final states (among others). No signal was found and
these bounds will apply to the case of a light hidden-Higgs/radion
in the present model. However, some work is required to translate these results to the present framework as the signal can also be generated via $e^+e^-\rightarrow h_1X_0$, followed by $h_1\rightarrow 2X_0$ (or $2h'$). It would be interesting to see if these results
allow one to improve the constraints on the kinetic-mixing with the hidden sector. Also,  a recent study on light hidden $U(1)$ factors
appeared in Ref.~\cite{Davoudiasl:2012ag}.
\appendix
\section{Phase Transitions and a Secluded Warped Throat\label{app:phase_trans}}
For a large-$N$ gauge theory with conformal dynamics near the confinement scale (dual to an RS model with negligible backreaction), demanding that the phase transition completes gives an upper bound on the number of colors $N$~\cite{Creminelli:2001th}. In this appendix we briefly discuss this bound, and how it can be alleviated in more general RS-like models.  First, note that one needs $(k/M_*)^3\lesssim [3\pi^3/(5\sqrt{5})]^{3/2}$ to ensure that the gravity description is reliable~\cite{Agashe:2007zd}. Combined with the expectation that $N^2\approx 16\pi^2(M_*/k)^3+1$ for RS models~\cite{Gubser:1999vj}, consistency requires $N\gtrsim 4$.\footnote{Ref.~\cite{Randall:2006py} uses a bound of $(k/M_*)^3\lesssim 16\pi^2$.} Demanding that the phase transition completes gives a bound on $N$ which can be estimated as follows~\cite{Kaplan:2006yi}: The bubble nucleation rate at temperature $\smash{T_c \sim \Lambda_\ir}$ is
\be
\Gamma_n \sim \Lambda_\ir^4 \exp(-c N^2) \, ,
\ee 
where $c=\mathcal{O}(1)$. The Hubble rate during this epoch is $H \sim N \Lambda_\ir^2/M_{Pl}$, and requiring that $\Gamma_n > H^4$ gives
\be
N^2 \ \lesssim\ \frac{4}{c}\ \log(M_{Pl}/\Lambda_\ir) \, .\label{bound_N}
\ee
Note that this becomes weaker with lower confinement scale $\Lambda_\ir$, though only logarithmically. For the specific value of $\Lambda_\ir= 1 \text{ GeV}$, one thus requires $N \lesssim 13$ to ensure that the phase transition completes, so $N$ cannot be arbitrarily large.

The upper bound on $N$ can be relaxed when the theory deviates from conformality (the geometry deviates from $AdS_5$) in the IR. Thus, including the leading order backreaction due to stabilization somewhat relaxes the bound~\cite{Konstandin:2010cd}. A more significant IR-deviation occurs for the warped RS-like models that arise in string theory~\cite{Klebanov:2000hb,Brummer:2005sh,Hebecker:2006bn}. These ``warped throats'' admit an RS-like description at low energies (below the effective 5D gravity scale), with the leading order correction to the RS geometry captured by a position-dependent curvature, $k\rightarrow k(z)$, which increases towards the IR~\cite{Brummer:2005sh}. The geometry is $AdS_5$-like in the UV and increasingly deviates from $AdS_5$ towards the IR. Thus, the dual theory has an energy-dependent number of colors, $N^2(z)\sim [M_*/k(z)]^3$, such that $N_{\uv}>N_{\ir}$, and is a cascading gauge theory~\cite{Klebanov:2000hb}. The demand that the phase transition completes puts a constraint on $N_{\ir}$~\cite{Hassanain:2007js}. However, because $N(z)$ decreases towards the IR, the effect of this constraint is less severe than in RS models (with constant $N_{\rs}$): A warped throat with UV value $N_{\uv}\sim N_{\rs}$ runs to an IR value $N_{\ir}< N_{\rs}$, thereby enabling a safe gravity description in the UV with $N_{\uv}\gg1$, and a successful phase transition in the IR with $N_{\ir}\sim\mathcal{O}(1)$. This allows the phase transition to be less-strongly first-order so the nucleation temperature can be close to the critical temperature, $T_n\simeq T_c$.

One could also implement our Secluded Dark Matter model on a warped throat (equivalently, with a cascading gauge theory). Although some quantitative differences would exist, the qualitative  picture would remain the same. The precise interaction strength between light KK-modes in the throat would be modified relative to the RS result, due to the fact that these modes are localized in the IR region, where the geometry deviates the most from $AdS_5$. However, the higher KK-modes, which are not localized in the IR, would retain many of their properties. These higher modes are dual to the CFT modes and thus the coupling of DM to the CFT should be similar for $M_{\dm}/\Lambda_{\ir}\gg1$. The coupling of DM and $\gamma'$ to the CFT will be governed by the large-$N$ regime near the UV brane and, provided the value of $N$ at freeze-out is on the order of (a given fixed value of) $N_{\rs}$, the results in the text should hold, up to $\mathcal{O}(1)$ corrections.
 \section{Dark Matter Abundance in the Presence of a CFT\label{app:DM_entopy}}
First-order phase transitions in the early universe can modify the annihilation cross section necessary to obtain the correct DM relic abundance~\cite{Wainwright:2009mq,Chung:2011hv} (for earlier work on late-time phase transitions see~\cite{Frieman:1991tu}). 
In what follows we extend our discussion of Section~\ref{sec:cos},  making use of
the analysis in Ref.~\cite{Megevand:2003tg}. We concentrate on the case of a phase transition without significant supercooling. In this case the `nucleation temperature', at which bubbles of the confined phase start growing faster than the expansion rate of the universe, is close to the critical temperature, $T_n \simeq T_c$. The latent heat released by these bubbles  
can therefore readily bring the temperature of the surrounding deconfined plasma back to $T_c$. This results in the coexistence of the confined and deconfined phases during the transition~\cite{Witten:1984rs}. The latent heat is gradually dissipated due to the expansion of the universe, so that regions containing the confined phase grow with time until they eventually occupy the entire space and the phase transition completes. As argued in~\cite{Witten:1984rs}, this process happens close to equilibrium.

Let us determine how much the universe expands during the phase transition. The deconfined phase at temperature $T_c$ (the common temperature of the deconfined and confined phases during the transition) has entropy density
\bea
s_{d}\, \sim\, (N^2 + g_\sm) \, T_c^3\ ,
\eea
since the CFT has $\mathcal{O}(N^2)$ degrees of freedom. On the other hand, the confined phase has $\mathcal{O}(1)$ degrees of freedom, so that
\bea
s_{c}\, \sim\, (\mathcal{O}(1) + g_\sm)\, T_c^3 \, \sim \, g_\sm \, T_c^3 \, .
\eea
Denoting the fraction of space in the
confined phase by $f$, the (averaged over space) entropy density is
\bea
s\, \sim\, (1-f) s_d + f s_c.
\eea
Since there is no significant departure from equilibrium, the comoving entropy is conserved to a good approximation (equivalently,
$s$ scales like $1/a^3$ with the expansion of the universe).\footnote{The bubble walls separating the confined and deconfined phase will eventually collide. This, and other effects, release some entropy and thus slightly increase the comoving entropy~\cite{Ignatius:1993qn}.} Thus,
the ratio of the scale factor $a_i$ when the phase transition starts
($f=0$), to the value $a_f$ when the transition completes ($f=1$) is
\bea
(a_i / a_f)^3\, =\, s_c / s_d\, \sim\, g_\sm / (N^2 + g_\sm).
\eea

In order to determine the
current DM abundance, one must evolve the number density at freeze-out to the
corresponding number density at our epoch. To achieve this one can divide the DM number density at
freeze-out by the total entropy density (i.e.~in both the CFT and SM sector) at that time. This quantity
stays constant over the subsequent expansion of the universe as both factors
scale like $1/a^3$ (because the comoving entropy density is conserved to good approximation). Multiplying this ratio by the
entropy density at our epoch (which is mostly carried by photons and is of
the order CMB temperature to the third power), one obtains the DM number
density at our epoch. As discussed in Section~\ref{sec:cos}, 
the correct abundance is then obtained if the DM has the annihilation cross section given in Eq.~\eqref{DMcs}.

Alternatively, one can divide the DM number density at freeze-out by the entropy
density of \emph{just the SM sector}, i.e.~$s_\sm \sim g_\sm T_f^3$, where $T_f$ is
the freeze-out temperature. In this case one must account for
the period (namely during the phase transition of the CFT)
during which the entropy density in the SM does not decrease under the
expansion of the universe (it instead remains constant at $s_\sm \sim g_\sm
T_c^3$). The expansion of the universe during this phase dilutes the DM
number density by an additional factor $(a_i / a_f)^3$.
Accounting for this dilution factor, and the fact that the universe expands faster during freeze-out by a factor of $\sqrt{(N^2+g_\sm)/g_{\sm}}$ due to the CFT (which follows from comparing the Hubble rate with and without the CFT), one requires the annihilation cross section
\bea
\langle \sigma v \rangle\ \approx  \sqrt{\frac{g_\sm}{N^2 + g_\sm}}\times   3\cdot 10^{-26}~\mathrm{cm}^3/\mathrm{s}
\eea
to obtain the right relic abundance in the presence of the CFT. This is in agreement with Eq.~\eqref{DMcs}.
Of course, both derivations of the requisite cross section are 
effectively the same, and are based on the (approximate) conservation of the comoving entropy density.

\end{document}